\documentclass[trackchanges, twocolumn]{aastex631}
\usepackage{amsmath}

\usepackage{makecell}

\usepackage{xcolor}

\graphicspath{{./}{figures/}}

\accepted{2021 November 30}
\submitjournal{ApJ}

\shorttitle{Outflow-galaxy connection in GRB hosts at $z > 2$}
\shortauthors{Gatkine et al.}

\begin{document}

\title{The CGM-GRB Study II: Outflow-Galaxy Connection at $z \sim 2-6$}


\correspondingauthor{Pradip Gatkine}
\email{pgatkine@caltech.edu}

\author[0000-0002-1955-2230]{Pradip Gatkine}
\altaffiliation{NASA Hubble Fellow}
\affil{Division of Physics, Mathematics and Astronomy, California Institute of Technology, Pasadena, CA 91125, USA}
\affil{Dept. of Astronomy, University of Maryland, College Park, MD 20742, USA}

\author[0000-0002-3158-6820]{Sylvain Veilleux}
\affiliation{Dept. of Astronomy, University of Maryland, College Park, MD 20742, USA}
\affiliation{Joint Space Science Institute, University of Maryland, College Park, MD 20742, USA}

\author[0000-0001-8472-1996]{Daniel Perley}
\affiliation{Astrophysics Research Institute, Liverpool John Moores University, IC2, Liverpool Science Park, 146 Brownlow Hill, Liverpool L3 5RF,UK}

\author{Joseph Durbak}
\affiliation{Dept. of Physics, University of Maryland, College Park, MD 20742, USA}

\author[0000-0001-6849-1270]{Simone Dichiara}
\affiliation{Dept. of Astronomy, University of Maryland, College Park, MD 20742, USA}
\affiliation{Astrophysics Science Division, NASA Goddard Space Flight Center, 8800 Greenbelt Rd, Greenbelt, MD 20771, USA}

\author[0000-0003-1673-970X]{S. Bradley Cenko}
\affiliation{Joint Space Science Institute, University of Maryland, College Park, MD 20742, USA}
\affiliation{Astrophysics Science Division, NASA Goddard Space Flight Center, 8800 Greenbelt Rd, Greenbelt, MD 20771, USA}

\author[0000-0002-1869-7817]{Eleonora Troja}
\affiliation{Dept. of Astronomy, University of Maryland, College Park, MD 20742, USA}
\affiliation{Astrophysics Science Division, NASA Goddard Space Flight Center, 8800 Greenbelt Rd, Greenbelt, MD 20771, USA}

\begin{abstract}
We use a sample of 27 GRBs at redshift $z=2-6$ to probe the  outflows in their respective host galaxies ($\mathrm{log(M_*/M_{\odot})}~\sim~9-11$) and search for possible relations between the outflow properties and those of the host galaxies such as $\mathrm{M_*}$, SFR, and specific SFR. First, we consider three outflow properties $-$ outflow column density ($\mathrm{N_{out}}$), maximum outflow velocity ($\mathrm{V_{max}}$), and normalized maximum velocity ($\mathrm{V_{norm}}$ = $\mathrm{V_{max}/V_{circ, halo}}$, where $\mathrm{V_{circ,halo}}$ is the halo circular velocity). We observe clear trends of $\mathrm{N_{out}}$ and $\mathrm{V_{max}}$ with increasing SFR in high-ion-traced outflows, with a stronger ($>~3\sigma$) $\mathrm{V_{max}}-$SFR correlation. We find that the estimated mass outflow rate and momentum flux of the high-ion outflows scale with SFR and can be supported by the momentum imparted by star formation (supernovae and stellar winds).  The kinematic correlations of high-ion-traced outflows with SFR are similar to those observed for star-forming galaxies at low redshifts.  

The correlations with SFR are weaker in low-ions. This, along with the lower detection fraction in low-ions, indicates that the outflow is primarily high-ion dominated. We also observe a strong ($>~3\sigma$) trend of normalized velocity ($\mathrm{V_{norm}}$) decreasing with halo mass and increasing with sSFR, suggesting that outflows from  low-mass halos and high-sSFR galaxies are most likely to escape and enrich the outer CGM and IGM with metals. By comparing the CGM-GRB stacks  with those of starbursts at $z\sim2$ and $z\sim0.1$,  we find that over a broad redshift range, the outflow strength strongly depends on the main-sequence offset at the respective redshifts rather than simply the SFR.
\end{abstract}


\section{Introduction} \label{sec:intro}

Galactic inflows and outflows shape the evolution of galaxies as well as enrich the circumgalactic medium (CGM) and intergalactic medium (IGM). The gas inflows fuel star formation while stellar winds, supernova (SN) explosions, and active galactic nuclei inject energy and metal-enriched matter (as well as entrained cold gas)  at  large  distances  into the interstellar medium (ISM) and CGM \citep{veilleux2005galactic, benson2010galaxy, booth2013interaction, tumlinson2017circumgalactic, rupke2018review,  veilleux2020cool}.  
The recycling flows from the CGM bring back the metal-enriched gas to refuel the star formation \citep{christensen2016n}. At the same time, removal of cold gas from the ISM can quench the star formation activity. Thus, galactic outflows regulate stellar buildup and are an important piece of the galactic feedback puzzle. AGN-driven outflows are thought to be the dominant feedback process in massive galaxies \citep{veilleux2005galactic, fabian2012observational, heckman2014coevolution, king2015powerful, nelson2019first} whereas SN-driven outflows are thought to be more important in low-mass, star-forming galaxies \citep{sharma2012supernovae}.



Supernova-driven outflows at high-redshift are important for the early enrichment of the CGM and IGM \citep{tumlinson2017circumgalactic, veilleux2020cool}. The low-mass star forming galaxies are of particular interest in this context since their outflows are most likely to escape their shallower potential wells. The relationship between the outflows and their host galaxies in the early universe holds the key to tune the models of galactic feedback and understand the history of galaxy growth and cosmic metal enrichment. 

Observations at high redshift ($z > 2$) using various techniques have shown the presence of ubiquitous outflows in star-forming galaxies. The prominent techniques include down-the-barrel absorption-line studies \citep{frye2002spectral, shapley2003rest, sugahara2017evolution, du2018redshift, rudie2019column}, outflows at larger radii using background quasar or galaxy sightlines \citep{steidel2010structure, lehner2014galactic, turner2014metal, rudie2019column}, quasar-quasar pairings \citep{hennawi2006quasars, prochaska2014quasars}, observing lensed galaxy spectra \citep{rigby2018magellan}, spatially-resolved spectroscopy in optical or radio \citep{harrison2012energetic, swinbank2015mapping, nielsen2020cgm, pizzati2020outflows}, and GRB afterglow sightlines \citep{fox2008high, gatkine2019cgm}. Galactic as well as cosmological zoom-in simulations provide the framework to understand the outflow mechanisms (for instance, \citealt{hirschmann2013effect, shen2013circumgalactic, muratov2015gusty, nelson2019first, mitchell2020galactic}). 
The high-$z$ outflow-galaxy relation and its evolution with redshift has recently been studied in 
\cite{sugahara2017evolution, sugahara2019fast}. 

However, the outflow-galaxy relation in low-mass  galaxies in the early universe remains poorly understood due to observational challenges. Two key challenges are: determining the redshift of the galaxy (in case of background QSO/galaxy sightlines) and obtaining high quality absorption spectra of these faint galaxies (for down-the-barrel technique). Apart from this, reliably removing the continuum spectrum of the background object can be a challenge. 

Use of GRB sightlines to probe the outflows and CGM of its host galaxy offers a promising solution to these problems. In \cite{gatkine2019cgm}, we described this method in detail. The main idea here is to use the bright GRB afterglow to probe the kinematics/outflows in the CGM of its host galaxy. GRB hosts at $z > 2$ are typically low-mass galaxies (log($\mathrm{M_*/M_{\odot}}$) $<$ 10.5), which makes them ideally suited for exploring the low-mass outflows which are difficult to probe using other techniques. The key advantages include: 1) clear identification of the host-galaxy redshift, 2) high signal-to-noise ratio (SNR) and high-resolution spectra due to the bright GRB afterglow, and 3) the featureless continuum of the GRB afterglow eliminates the problem of continuum subtraction.



In this paper, we use the CGM-GRB sample compiled in \cite{gatkine2019cgm} to explore the correlations between outflow and galaxy properties. The CGM-GRB sample consists of 27 GRBs at $z \sim 2-6$ with high SNR (median SNR $\sim$ 10) and high-resolution ($\delta v < 50$ km s$^{-1}$) spectra. Multi-component Voigt-profiles were fit to the absorption spectra of various high- and low-ion species (including C IV, Si IV, Si II, Fe II, and O VI). The CGM kinematics of this sample were studied in \cite{gatkine2019cgm}. In this paper, we report the observations of their host galaxies in the optical and near-IR to estimate their star formation rate (SFR) and stellar mass ($\mathrm{M_*}$). These observations and their analyses are described in Section \ref{sec:data}. We then discuss the techniques used for visualizing and inferring correlations in Section \ref{sec:analysis}. The key correlations between outflow properties and galaxy properties such as $\mathrm{M_{*}}$, SFR, specific star formation rate (sSFR = SFR/$\mathrm{M_*}$), and halo mass  are detailed in Section \ref{sec:outflow_correlations}. Finally, the implications of our results are discussed in Section \ref{sec:discussion}.  

Throughout this paper, we use the following model of cosmology: $H_0 = $ 70 $\mathrm{km ~s^{-1}Mpc^{-1}}$, $\Omega_{M} = 0.3$, $\Omega_{\Lambda} = 0.7$


\section{Observations and Methods} \label{sec:data}

As described earlier, we measure the galaxy properties in the CGM-GRB sample. The sample is selected strictly on the criterion of availability of a high-resolution ($\delta v$ $<$ 50 km s$^{-1}$) and high-SNR (SNR $>$ 5) afterglow spectrum. No cuts are made to the sample based on galaxy properties. The redshift distribution of the sample is shown in the first panel of Fig. \ref{fig:Sample_properties}.

\subsection{Optical Photometry}
\label{subsec:optical_photometry}

We performed optical photometry of previously unpublished or unobserved GRB hosts in the CGM-GRB sample. We observed GRB hosts using the 4.3-meter Lowell Discovery Telescope (LDT). We also obtained deep archival imaging of two GRB hosts using the FORS instrument on the Very Large Telescope (VLT) and one each using HST WFC3 (program ID 15644), the Kilo-Degree Survey (KiDS) \citep{kuijken2019fourth}, and PanSTARSS survey \citep{flewelling2016pan}. We consider a GRB host as detected if the offset of the potential host and the GRB location is within 1\arcsec. The probability of a chance alignment of a galaxy brighter than the typical depth in our observations ($i$-band $\sim$ 24.7 AB mag) within 1\arcsec is approximately 0.01 (see Fig. 6 in \citealt{beckwith2006hubble}).  
At $z \sim$ 3, 1\arcsec\ roughly corresponds to 7.5 kpc. From previous HST observations of other GRB host samples at  \citep{bloom2002observed, fruchter2006long, lyman2017host}, more than $90\%$ of the GRBs occur within this offset from their host galaxies. All the GRBs are localized with a $< 0.5\arcsec$ precision. The resulting magnitudes are further corrected for Milky Way Galactic extinction using the dust maps of \cite{schlafly2011measuring} and the extinction law with $R_{V}$ = 3.1 from \cite{cardelli1989relationship}. The photometry results are presented in Table \ref{tab:GRB_Host_Photometry}.  

The LDT imaging was performed using the Large Monolithic Imager (LMI, \cite{massey2013big}). The LMI data was detrended with a custom python-based pipeline \citep{toy2016exploring}. 
Individual fields were astrometrically aligned and co-added using {\fontfamily{qcr}\selectfont SCAMP} and {\fontfamily{qcr}\selectfont SWARP} respectively. The aperture photometry of the co-added images was performed using {\fontfamily{qcr}\selectfont Sextractor} with an aperture radius of $\sim$1.5\arcsec, which is typical of the average seeing in our observations. The magnitudes were calibrated against the SDSS \citep{alam2015eleventh} and GAIA catalogs \citep{evans2018gaia}. Conversion of GAIA magnitudes to Sloan magnitudes was performed using the conversion tables provided in GAIA data release 2 \citep{brown2018gaia}. 

The FORS data was flat-fielded using the ESO pipeline {\fontfamily{qcr}\selectfont ESOreflex}  \citep{freudling2013automated} and was further aligned, co-added, and calibrated as described above. PanSTARSS and KiDS surveys provide reduced, stacked, and zero-point calibrated images, which were used to determine the science magnitudes/upper limits. The HST photometry was performed using archived drizzled and calibrated images and the AB magnitude was derived using the provided zero point. A 1\arcsec\ aperture was used for HST images given the diffraction-limited imaging.

\begin{deluxetable}{ccccc}[t]
\tablecaption{Summary of new observations \label{tab:GRB_Host_Photometry}}
\tablehead{
\colhead{GRB} &
\colhead{$z^{c}$} &
\colhead{Tel./Instr.} &
\colhead{Filter} & 
\colhead{AB Mag} 
}

\startdata
000926A$^a$ & 2.0377 &  {\em Spitzer}/IRAC & 3.6 $\mu m$ & 25.2 $\pm$ 0.15 \\
021004$^a$ & 2.3281 &  {\em Spitzer}/IRAC & 3.6 $\mu m$ & 24.22 $\pm$ 0.18 \\
071031$^b$ & 2.6912 &  {\em Spitzer}/IRAC & 3.6 $\mu m$ & $>$ 25.3 \\
080310$^b$ & 2.4274 & {\em Spitzer}/IRAC & 3.6 $\mu m$ & 23.74 $\pm$ 0.24 \\
090926A & 2.106 & \makecell{{\em Spitzer}/IRAC\\ VLT/FOR2} & \makecell{3.6 $\mu m$\\ R$_\mathrm{Special}$} & \makecell{22.96 $\pm$ 0.05\\ 23.9 $\pm$ 0.1} \\
111008A$^b$ & 4.989 & \makecell{ {\em Spitzer}/IRAC\\ HST/WFC3} & \makecell{3.6 $\mu m$\\ F110W} & \makecell{24.73 $\pm$ 0.3\\ 25.5 $\pm$ 0.07} \\
120327A & 2.813 & LDT/LMI & SL-r & 24.9 $\pm$ 0.2\\
130606A & 5.911 & {\em Spitzer}/IRAC & 3.6 $\mu m$ & 24.91 $\pm$ 0.25 \\
130610A & 2.091 & \makecell{{\em Spitzer}/IRAC\\ LDT/LMI} & \makecell{3.6 $\mu m$\\ SL-r} & \makecell{23.46 $\pm$ 0.05\\ 23.7 $\pm$ 0.1} \\
141028A & 2.333 & \makecell{{\em Spitzer}/IRAC\\ LDT/LMI} & \makecell{3.6 $\mu m$\\ SL-r} & \makecell{$>$ 25.1 \\ $>$ 25.8} \\
141109A & 2.993 & \makecell{{\em Spitzer}/IRAC\\ LDT/LMI} & \makecell{3.6 $\mu m$\\ SL-i} & \makecell{23.4 $\pm$ 0.1\\ 24.1} \\
151021A & 2.329 & \makecell{{\em Spitzer}/IRAC\\ KiDS Survey} & \makecell{3.6 $\mu m$\\ SL-r} & \makecell{$>$ 25.7\\ 24.4 $\pm$ 0.2} \\
151027B & 4.0633 & \makecell{{\em Spitzer}/IRAC\\ LDT/LMI} & \makecell{3.6 $\mu m$\\ SL-r\\ SL-i} & \makecell{$>$ 22.66\\ $>$ 24.3\\ 24.8 $\pm$ 0.4} \\
160203A & 3.518 & \makecell{{\em Spitzer}/IRAC\\ PanSTARRS} & \makecell{3.6 $\mu m$\\ PS1-i} & \makecell{21.74 $\pm$ 0.02\\ $>$ 22.7} \\
161023A & 2.709 & \makecell{{\em Spitzer}/IRAC\\ VLT/FORS2} & \makecell{3.6 $\mu m$\\ R$_\mathrm{Special}$} & \makecell{$>$ 25.9\\ $>$ 25.7} \\
170202A & 3.645 & LDT/LMI &  \makecell{SL-r\\ SL-i} &  \makecell{$>$ 25.4\\ $>$ 23.4} \\
\enddata

\tablenotetext{a}{{\em Spitzer} Prog ID 40599, PI : R. Chary.}
\tablenotetext{b}{{\em Spitzer} Prog ID 80054, PI : E. Berger. All other {\em Spitzer} observations are taken from {\em Spitzer} Prog IDs 11116, 13104, 90062  PI: D. Perley.} 
\tablenotetext{c}{Redshifts taken from \cite{gatkine2019cgm}}

\end{deluxetable}

\begin{figure*}
\centering
\includegraphics[width=0.9\textwidth]{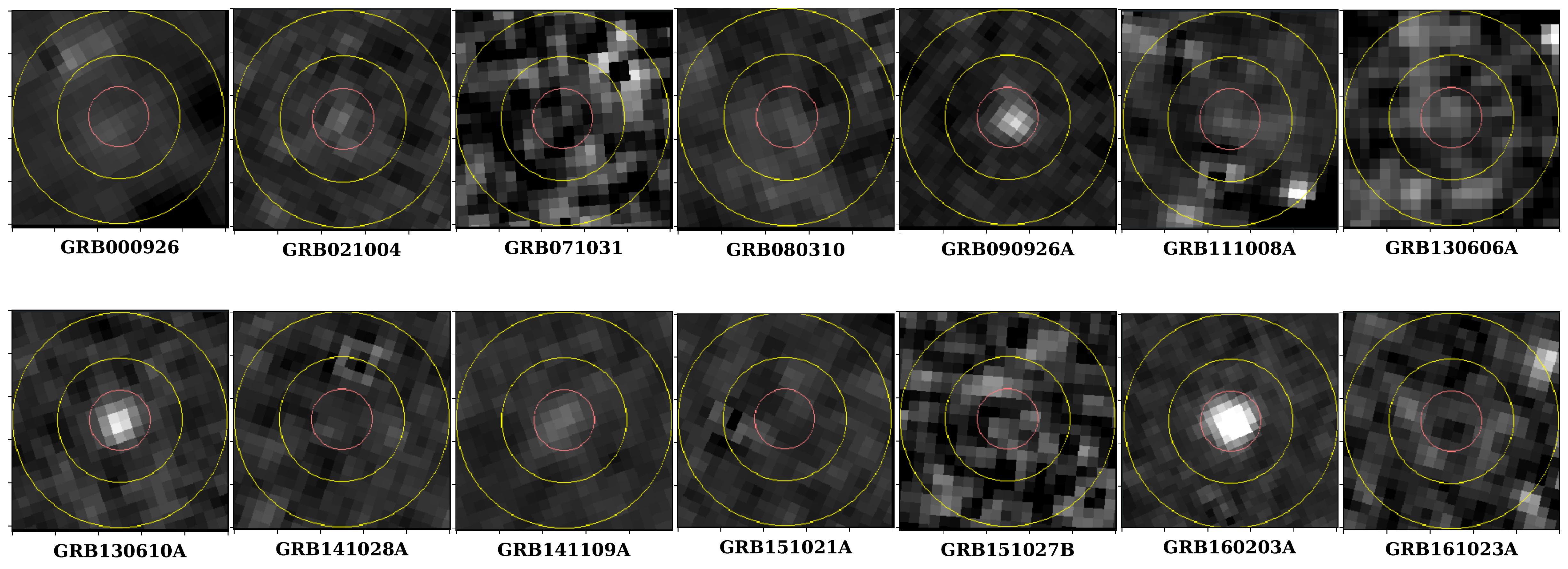}
\figcaption{\label{fig:Spitzer_postage} 
 Contamination-subtracted images of GRB fields from {\em Spitzer}-IRAC in 3.6 $\mu m$ band. Each thumbnail is 8" $\times$ 8" in size. The central red circle is the 1.8\arcsec aperture used to define the source flux and the outer annulus is used to define the background flux. The circle is centered on the best-known position of the GRB or of the detected host galaxy. References for GRB positions: 000926 \citep{fynbo2001optical},  021004 \citep{henden2002grb021004},  071031 \citep{kruhler2009correlated},  080310 \citep{littlejohns2012origin},   111008A \citep{bolmer2018dust},  130606A \citep{castro2013grb},  141109A \citep{xu2014x},  151021A \citep{mccauley2015grb},  151027B \citep{greiner2018grb},   161023A \citep{de2018x}.}
\end{figure*}

\subsection{Spitzer IRAC Photometry}
\label{subsec:IR_photometry}
We obtained deep archival imaging of GRB hosts using {\em Spitzer} Infrared Array Camera (IRAC) channel 1 ($3.6 \mu m$). Out of a total of 27 GRB hosts, we present new {\em Spitzer} IRAC photometry of 14 hosts in this paper and 11 were previously published as a part of the SHOALS survey \citep{perley2016swift_2}. The remaining two GRBs remained unobserved by the end of {\em Spitzer} mission. The newly presented data have been collected as a part of various previous programs which are summarized in Table \ref{tab:GRB_Host_Photometry}.    

By analyzing the new data the same way as \cite{perley2016swift_2}, we ensure procedural consistency with the previously published data. The reduction and photometry method is described in detail in \cite{perley2016swift_2}. Here, we briefly summarize the key points. We acquired the Level-2 PBCD (Post-Basic Calibrated Data) from the {\em Spitzer} Legacy Archive. We use the default astrometry provided with the Level-2 products (with an accuracy of 0.3\arcsec). Due to the large PSF of {\em Spitzer} IRAC ($\sim$ 1.8\arcsec\ at 3.6 $\mu m$), source confusion and flux contamination from neighboring sources is an important issue. We compare each IRAC  image with deep ground-based optical images (as described in Section \ref{subsec:optical_photometry}) to identify the primary source and any neighboring contaminants. We used the {\fontfamily{qcr}\selectfont galfit} tool \citep{peng2002detailed} over several iterations to model the sources (using the PSF and PRF files provided in {\em Spitzer} documentation\footnote{https://irsa.ipac.caltech.edu/data/{\em Spitzer}/docs/irac/calibrationfiles/psfprf/}) and subtract the neighboring sources which may contaminate the host or sky background regions. The subtracted image is then used for performing aperture photometry. 

We implemented the IRAC handbook recommendations for aperture photometry using a custom IDL wrapper around the {\fontfamily{qcr}\selectfont aper} procedure in the Astronomy User's Library\footnote{https://idlastro.gsfc.nasa.gov/} (see \cite{perley2016swift_2} for details). For aperture photometry, we place a 1.8\arcsec\ aperture on the host galaxy location (guided by deep optical imaging) and a sky annulus with an inner radius of 3.6\arcsec\ and outer radius of 6\arcsec. The source aperture and sky annulus are marked in red and yellow respectively in Fig. \ref{fig:Spitzer_postage}. In the case of optical detection and IR non-detection, we specify a 2-$\sigma$ limit. However, in the case of optical as well as IR non-detection, we evaluate a 3-$\sigma$ upper limit to account for the uncertainty (typically $<$ 1\arcsec) in the GRB host location.   

\subsection{Stellar Mass}
We use the {\em Spitzer} IRAC 3.6$\mu m$ photometry to infer the stellar mass of galaxies in our sample. At $z$ $\sim$ $2-6$, {\em Spitzer} IRAC measures the rest-frame optical light (beyond the Balmer break) from long-lived stars in the host galaxies. Here we follow the methodology used in \cite{perley2016swift_2} to derive the stellar masses. SED fitting is a more accurate method to estimate $\mathrm{M_{*}}$ (by breaking the degeneracy between age and extinction). However, this requires extensive, ultra-deep optical observations of faint GRB hosts in multiple filters, which is resource intensive. Instead, we use {\em Spitzer} single-band (3.6 $\mu m$) photometry, which can still provide a reasonable estimate of stellar mass, particularly for galaxies at $z > 2$. 

We calculate the absolute magnitude at $\lambda_{rest} = 3.6 \mu m/(1+z)$ as $\mathrm{M_{AB}}$ = $m_{\mathrm{AB, 3.6 \mu m}} -$ DM + 2.5log(1+$z$), where DM is distance modulus. In \cite{perley2016swift_2}, a grid of model galaxy SEDs is constructed for an array of redshifts ($z \sim 0-10$) and each decade in $\mathrm{M_{*}}$ ($10^8$ to $10^{11} \mathrm{M_{\odot}}$) by summing \cite{bruzual2003stellar} galaxy SED templates (using \cite{chabrier2003galactic} initial mass function). The models also incorporate a modest dust attenuation to validate the single-band stellar mass conversion function against the more accurate SED-fit (optical $+$ {\em Spitzer} multiband) stellar masses in the MODS \citep{kajisawa2009moircs} and UltraVISTA samples \citep{caputi2015Spitzer}. We then evaluate the stellar mass by interpolating on the $\mathrm{M_{*}}$, redshift, and AB magnitude grid (see \cite{perley2016swift_2} for more details). While the single-band method suffers from uncertainties associated with various model assumptions such as the IMF, dust extinction ($\mathrm{A_{V}}$), and star formation history, this method is consistent with the masses obtained from SED fitting at the $\sim$ 0.3 dex level. Further, by using the same method throughout our sample, we ensure that the correlations derived here are on an equal footing. The $\mathrm{M_{*}}$ of our GRB hosts are summarized in Table \ref{tab:GRB_SFR}.

\begin{figure*}
\centering
\includegraphics[width=0.95\textwidth]{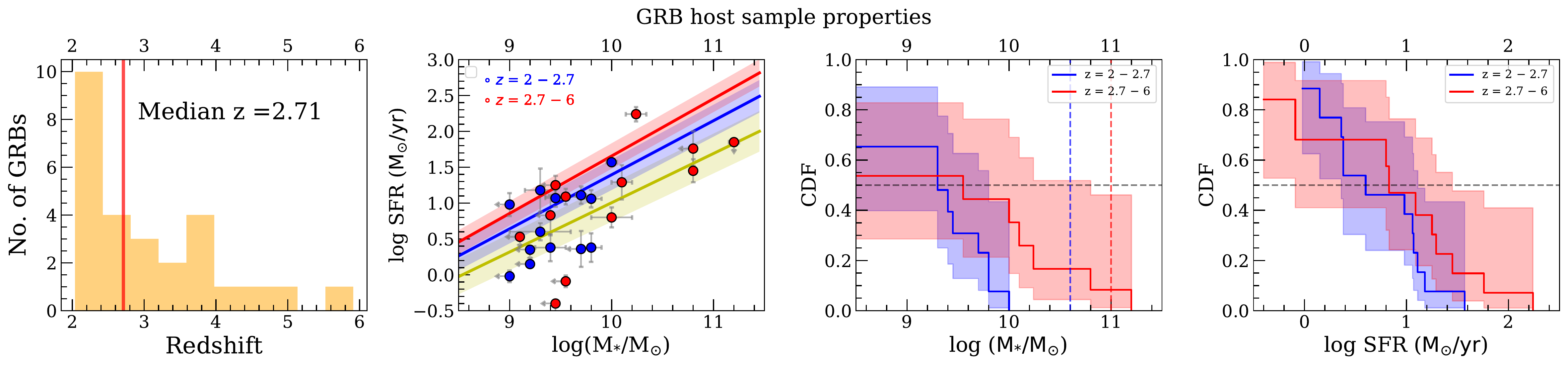}
\figcaption{\label{fig:Sample_properties} 
Properties of the CGM-GRB sample. Panel 1: The redshift distribution of the sample. Panel 2: The SFR $vs$ $\mathrm{M_*}$ of the GRB host galaxies in our sample. The lines show the main sequence curves (yellow: $z = 1$, blue: $z = 2$, red: $z = 4$) as described in \cite{speagle2014highly}. Panel 3: The cumulative distriution of the stellar mass in the CGM-GRB sample. The spread shows 95\% confidence interval around the value by incorporating any upper limits. The dotted vertical lines show the value of characteristic mass, $\mathrm{M^*}$ in the mass function (written as a Schechter function) at the respective redshifts. The horizontal line shows the  median (i.e. CDF = 0.5). Panel 4:  Same as panel 3, for SFR. }
\end{figure*}

\subsection{Dust Correction \label{subsec: Dust_corr}}
The ultraviolet (UV) dust extinction of the host galaxies needs to be estimated to convert the absolute magnitudes into intrinsic rest-frame UV luminosities. Following \cite{greiner2015gamma}, we perform the dust correction using empirical correlations of the spectral index of  the UV continuum $\beta$ (where $f_{\lambda} = \lambda^{\beta}$), rest-frame absolute UV magnitude at $\lambda_{rest}$ = 1600 $\mathrm{\AA}$ ($\mathrm{M_{UV}}$), and the dust extinction at rest-frame $1600 \mathrm{\AA}$ ($A_{1600}$). Here we assume that GRB hosts at high redshift follow a power law SED ($f_{\lambda} = \lambda^{\beta}$) in the UV (redward of Ly$\alpha$) and the same correlations as the extensive high-$z$ ($2.5 - 6$) star forming galaxy sample of $> 4000$ galaxies from HST HUDF and CANDELS surveys studied in \cite{bouwens2009uv, bouwens2014uv}. They derive the following empirical relation for star forming galaxies at $\left \langle z \right \rangle$ = 3.8: 
\begin{equation} \label{eqn:beta}
    \beta = -1.85 - 0.11(\mathrm{M_{UV}} + 19.5)
\end{equation}

The uncertainties on the numerical coefficients here are small ($-1.85\pm{0.06}$ and $-0.11\pm{0.01}$). Then, we iteratively solve for $\mathrm{M_{UV}}$ and $\beta$. The typical $\beta$ for the high-$z$ star forming sample in \cite{bouwens2009uv, bouwens2014uv} is $\beta$ $\sim$ $-2$. In equation \ref{eqn:beta}, this corresponds to $\mathrm{M_{UV} = -18.1}$. Hence, we use  $\beta$ $=$ $-2$ for our weaker upper limits (where $\mathrm{M_{UV, lim} > -18.1}$), where the value of $\beta$ is more uncertain. For stronger upper limits (i.e. $\mathrm{M_{UV, lim} < -18.1}$, we use the $\beta$ corresponding to the limit. Finally, the $A_{1600}$ is evaluated using the following relation from \cite{meurer1999dust}: 

\begin{equation} \label{eqn:beta_2}
    A_{1600} = 4.43 ~\mathrm{mag} + 1.99\beta
\end{equation}

This dust-correction method is described in detail in \cite{greiner2015gamma}. 

\subsection{Star formation rate}
We use single-band photometry in the rest-frame UV to calculate UV-based SFR. To compute the SFR from dust-corrected UV luminosity ($L_{UV, corr}$), we follow the relations described in \cite{savaglio2009galaxy} where they simultaneously compare the emission-line and dust-corrected UV luminosities of GRB hosts to derive the conversion factor between dust-corrected UV luminosity and SFR. We use the $A_{1600}$ and $\mathrm{M_{UV}}$ values calculated in Section \ref{subsec: Dust_corr} to compute $L_{UV, corr}$. The SFR is then calculated as follows: 

\begin{equation}
    \mathrm{SFR_{1500}} = 1.62 ~\mathrm{M_{\odot} yr^{-1}} \times  \frac{L_{\mathrm{1500, corr}}}{\mathrm{10^{40} ~erg~s^{-1}~\AA^{-1}}}
\end{equation}

As a validation step, we compare the $\mathrm{A_{1600}}$ evaluated using the $\beta$ method with that using the afterglow-derived $\mathrm{A_{V}}$ (assuming an SMC extinction law). The resulting SFRs derived using the two methods are consistent with each other within a factor of two except for GRBs 130408A and 080810 where the afterglow $\mathrm{A_{V}}$ is larger, leading to a higher SFR estimate (for the afterglow $\mathrm{A_{V}}$ method) by a factor of 3. Nonetheless, it is important to note that the afterglow-derived extinction corresponds to a single sightline, while the extinction derived using the $\beta$ method is an average value for the host.    The star formation rates of our GRB hosts are summarized in Table \ref{tab:GRB_SFR}.  

For GRBs 071031, 080804, and 120815, photometric observations are either unavailable or too shallow. In the cases of GRBs 080804 and 120815, we have used H$\alpha$ emission-line-based SFRs from \cite{kruhler2011seds} since they are more robust compared to UV-luminosity. For GRB 071031, we use the Ly$\alpha$-based SFR from \cite{milvang2012optically}. While less robust, this measurement is consistent with the upper limit of 3 $\mathrm{M_{\odot} yr^{-1}}$ from an archival HST WFC3 (F160W filter) observation.


Note however that our sample naturally has low line-of-sight dust extinction compared to the general GRB host population since we only select the afterglows that are bright enough for high-resolution rest-frame UV spectroscopy. While there may be a systematic bias in the dust correction, we have used the same SFR-tracer and analysis procedure for the entire sample (except GRBs 071031, 080804, and 120815), thus minimizing any relative bias. Our sample may contain a small number of heavily dust enshrouded galaxies, for which we may underestimate the SFR. However, we have minimized this possibility by ruling out heavy dust obscuration in 4 massive GRB hosts in our sample (where the probability of heavy dust obscuration is high) by using deep VLA observations \citep{gatkine2020new} and hence, the typical dust corrections described here can be used for estimating their star formation rates. These GRBs are marked with asterisk in Table \ref{tab:GRB_SFR}.

\begin{deluxetable*}{ccccccccc}[t]
\tabletypesize{\scriptsize}
\tablecaption{Summary of GRB host properties in the CGM-GRB sample \label{tab:GRB_SFR}}
\tablehead{
\colhead{GRB} &
\colhead{$z$} &
\colhead{log(N$_\mathrm{HI}$)$^{a}$} &
\colhead{A$_\mathrm{V}$\tablenotemark{b}} &
\colhead{M$_\mathrm{3.6/(1+z)}$} &
\colhead{log($\mathrm{M_{*}/M_{\odot}}$)} & 
\colhead{M$_\mathrm{UV}$} &
\colhead{\makecell{SFR\\ ($\mathrm{M_{\odot} yr^{-1}}$)}} & 
\colhead{References} 
}

\startdata
\vspace{-2ex}
000926 & 2.0385 & 21.3 $\pm$ 0.25 & 0.15 & $-19.6$ & 9.3$\pm$0.3 & $-19.5$ & 4.0$^{+1.3}_{-1.0}$ & \makecell{\vspace{-0.5ex} \cite{castro2003keck}\\ \cite{chen2009high}} \\ \vspace{-2ex}
021004$^*$ & 2.3281 & 19.0 $\pm$ 0.2 & 0.2 & $-20.9$ & 9.5$\pm$0.1 & $-21.4$ & 11.8$^{+3.7}_{-2.8}$ &  \makecell{\vspace{-0.5ex} \cite{fiore2005flash}\\ \cite{fynbo2005afterglow} } \\ \vspace{-2ex}
050730 & 3.9672 & 2.1 $\pm$ 0.1 & 0.12 & $>$ $-20.5$ & $<$ 9.46\tablenotemark{c} & $-18.1$ & 0.8$^{+0.2}_{-0.1}$ &  \makecell{\vspace{-0.5ex} \cite{d2007uves}\\ \cite{toy2016exploring}}\\ \vspace{-2ex}
050820A & 2.6137 & 21.1 $\pm$ 0.1 & 0.08\tablenotemark{d} & $-20.42$ & 9.4$\pm$0.15\tablenotemark{d} & $-19.1$ & 2.4$^{+1.3}_{-0.9}$ &  \makecell{\vspace{-0.5ex} \cite{prochaska2007interstellar}\\ \cite{chen2009high}}\\ \vspace{-2ex}
050922C & 2.1996 & 21.55 $\pm$ 0.1 & 0.10 & $-19.6$ & $<$ 9.0\tablenotemark{d} & $>$ $-18.3$ & $<$ 1.0 &  \makecell{\vspace{-0.5ex} \cite{prochaska2008survey}\\\cite{covino2013dust}} \\ \vspace{-2ex}
060607A & 3.0738 & 16.95 $\pm$ 0.03 & 0.08 & $>$ $-20.52$ & $<$ 9.4\tablenotemark{d} & $>$ $-17.5$ & $<$ 0.4 &  \makecell{\vspace{-0.5ex} \cite{prochaska2008survey}\\ \cite{schady2012dust}} \\ \vspace{-2ex}
071031 & 2.6912 & 22.15 $\pm$ 0.05 & 0.14 & $>$ $-20.1$  & $<$ 9.2 & $-$ & 1.4$^{+0.3}_{-0.3}$\tablenotemark{e} &  \makecell{\vspace{-0.5ex} \cite{fox2008high}\\ \cite{li2018large}} \\  \vspace{-2ex}
080310$^*$ & 2.4274 & 18.7 $\pm$ 0.1 & 0.10 & $-21.3$ & 9.8$\pm$0.1\tablenotemark{d} & $-19.0$ & 2.4$^{+1.4}_{-0.9}$ &  \makecell{\vspace{-0.5ex} \cite{fox2008high}\\ \cite{perley2009host}} \\ \vspace{-2ex}
080804 & 2.205 & 21.3 $\pm$ 0.1 & 0.17 & $-20.2$ & 9.3$\pm$0.15\tablenotemark{d} & $-$ & 15.1$^{+20}_{-7}$\tablenotemark{e} &  \makecell{\vspace{-0.5ex} \cite{fynbo2009low}\\ \cite{toy2016exploring}}\\ \vspace{-2ex}
080810$^*$ & 3.351 & 17.5 $\pm$ 0.15 & 0.40 & $-22.15$ & 10.24$\pm$0.1\tablenotemark{d}  & $-22.9$ & 173$^{+45}_{-36}$  &  \makecell{\vspace{-0.5ex} \cite{page2009multiwavelength}\\ \cite{wiseman2017evolution}} \\ \vspace{-2ex}
090926A & 2.106 & 21.73 $\pm$ 0.07 & $<$ 0.04 & $-21.9$ & 9.8$\pm$0.1 & $-20.5$ &  11.6$^{+3.7}_{-2.8}$&  \makecell{\vspace{-0.5ex} \cite{d2010vlt}\\ \cite{zafar2018x}} \\  \vspace{-2ex}
100219A & 4.665 & 21.13 $\pm$ 0.12 & 0.13 & $>$ $-20.4$ & $<$ 9.4\tablenotemark{d} & $-20.0$ & 6.7$^{+5.5}_{-3.2}$ &  \makecell{\vspace{-0.5ex} \cite{thone2012grb}\\ \cite{toy2016exploring}} \\  \vspace{-2ex}
111008A & 4.989 & 22.3 $\pm$ 0.06  & 0.12 & $-20.9$ & 9.5$\pm$0.2 & $-20.5$ & 12.3$^{+3.5}_{-2.7}$ &  \makecell{\vspace{-0.5ex} \cite{sparre2014metallicity}\\ \cite{zafar2018x}}\\ \vspace{-2ex}
120327A & 2.813 & 22.01 $\pm$ 0.09 & $<$ 0.03 & $-23.2$ & 10.8$\pm$0.1 & $-21.2$ & 28.1$^{+12.5}_{-8.7}$ &  \makecell{\vspace{-0.5ex} \cite{d2014vlt}\\ \cite{heintz2019new}} \\  \vspace{-2ex}
120815A & 2.358 & 21.95 $\pm$ 0.1 & 0.19 $\pm$ 0.04 & $>$ $-21.2$ & $<$ 9.7\tablenotemark{d} & $-$ &  2.3$^{+2}_{-1}$\tablenotemark{e} &  \makecell{\vspace{-0.5ex} \cite{kruhler2015grb}\\ \cite{zafar2018x}} \\  \vspace{-2ex}
120909A & 3.929 & 21.20 $\pm$ 0.10 & 0.16 $\pm$ 0.04 & $>$ $-20.2$  & $<$ 9.5\tablenotemark{d} & $-20.8$ & 17.9$^{+6.2}_{-4.6}$ &  \makecell{\vspace{-0.5ex} \cite{cucchiara2015unveiling}\\ \cite{heintz2019new}} \\  \vspace{-2ex}
121024A$^*$ & 2.298 &  21.50 $\pm$ 0.10 & 0.56\tablenotemark{d} & $-21.8$ & 10.15$\pm$0.15 & $-21.7$ & 37$^{+20}_{-15}$ &  \makecell{\vspace{-0.5ex} \cite{friis2015warm}\\ \cite{toy2016exploring}} \\  \vspace{-2ex}
130408A & 3.757 & 21.70 $\pm$ 0.10 & 0.2 & $-$  & $-$ & $>$ $-21.1$ & $<$ 13.4 &  \makecell{\cite{zafar2018x}} \\  \vspace{-2ex}
130606A	& 5.911 &  19.93 $\pm$ 0.2 & $<$ 0.07 & $-21.8$  & 10.0$\pm$0.2 & $-19.9$ & 6.3$^{+2.4}_{-1.7}$ &  \makecell{\vspace{-0.5ex} \cite{hartoog2015vlt}\\ \cite{zafar2018x}} \\  \vspace{-2ex}
130610A & 2.091 & $-$ & 0.01 & $-21.3$  & 9.7$\pm$0.05 & $-20.6$ & 13$^{+4.1}_{-3.1}$ &  \makecell{\vspace{-0.5ex} \cite{2013GCN.14848....1S}\\ \cite{littlejohns2015detailed}} \\  \vspace{-2ex} 
141028A & 2.333 & 20.60 $\pm$ 0.15 & 0.13 & $>$ $-20.0$  & $<$ 9.2 & $>$ $-19.2$ & $<$ 2.3 &  \makecell{\cite{wiseman2017evolution}} \\ \vspace{-2ex} 
141109A & 2.993 & 22.10 $\pm$ 0.10 & 0.11 & $-22.1$ & 10.1$\pm$0.1 & $-20.9$ & 19.7$^{+14}_{-8}$ &  \makecell{\vspace{-0.5ex} \cite{heintz2018highly}\\ \cite{heintz2019new}}\\ \vspace{-2ex}
151021A & 2.329 & 22.3 $\pm$ 0.2 & 0.2 & $>$ $-19.4$ & $<$ 9.0 & $-20.3$ & 9.6$^{+4.3}_{-3}$ &  \makecell{\cite{heintz2018highly}} \\  \vspace{-2ex}
151027B & 4.0633 &  20.5 $\pm$ 0.2 & $<$ 0.12 & $-23.45$  & $<$ 10.8 & $-21.9$ & 58$^{+40}_{-24}$ &  \makecell{\vspace{-0.5ex} \cite{heintz2018highly}\\ \cite{zafar2018x}} \\ \vspace{-2ex}
160203A & 3.518 & 21.75 $\pm$ 0.10 & $<$ 0.1 & $-24.2$ &  11.2$\pm$0.05 & $>$ $-22.9$ & $<$ 71 & \makecell{\cite{heintz2018highly}} \\ \vspace{-2ex}
161023A & 2.709 & 20.96 $\pm$ 0.05 & 0.09 & $>$ $-19.5$ & $<$ 9.1 & $>$ $-19.6$ & $<$ 3.4 &  \makecell{\vspace{-0.5ex} \cite{heintz2018highly}\\  \cite{de2018x}} \\  \vspace{-2ex}
170202A & 3.645 & 21.55 $\pm$ 0.10 & $<$0.12 & $-$ & $-$ & $>$ $-21.0$ & $<$ 11.5 &  \makecell{\vspace{-0.5ex} \cite{selsing2019x}\\ \cite{zafar2018x}} \\ \vspace{-2ex}
\enddata
\vspace{1ex}
Column Descriptions: M$_\mathrm{3.6/(1+z)}$: AB magnitude in rest-frame optical/NIR from {\em Spitzer} data; M$_\mathrm{UV}$: Absolute magnitude at $\lambda_{rest}$ = 1600\AA ; SFR: in units of $\mathrm{M_{\odot}/yr}$
\tablenotetext{a}{Neutral hydrogen column densities (in cm$^{-2}$) measured from the damped Ly-$\alpha$ absorption, unless noted otherwise. \vspace{-0.5ex}}
\tablenotetext{b}{Extragalactic dust extinction in magnitude, derived assuming SMC extinction law \citep{gordon2003quantitative} \vspace{-0.5ex}}
\tablenotetext{c}{Derived Using host galaxy SED \vspace{-0.5ex}}
\tablenotetext{d}{From \cite{perley2016swift_2} \vspace{-0.5ex}}
\tablenotetext{e}{071031: SFR using Ly$\alpha$ \cite{milvang2012optically}, 080804: SFR using H$\alpha$ \cite{kruhler2015grb}, 120815: SFR using H$\alpha$ \cite{kruhler2015grb}.\\
\noindent $^*$ GRBs with deep VLA observations from \cite{gatkine2020new}
\vspace{-0.5ex}}
\end{deluxetable*}

\section{Sample properties and analysis} \label{sec:analysis}
\subsection{Comparison with star formation main sequence}
Figure \ref{fig:Sample_properties} shows the distribution of $\mathrm{M_{*}}$,  SFR, and $z$ of the CGM-GRB sample. We compare the relative position of our sample with respect to the star forming main sequence at $z =$ 2 and 4. The star forming main sequence and its scatter is computed using Equation (28) in \cite{speagle2014highly}. The key characteristics of our sample in terms of galaxy properties are summarized below.\\ \vspace{-2ex}

\noindent 
1. We divide the sample in two groups $-$  $z1$: 2-2.7 and $z2$: 2.7-5.9 $-$ which have equal number of objects and roughly equal cosmological timescale (1 and 1.4 Gyrs). We highlight that there is no significant difference in the two groups in terms of SFR distribution. On the other hand, the host-galaxy stellar mass distribution of the high-$z$ group is biased towards higher masses as shown in Fig. \ref{fig:Sample_properties} (panel 3). However, note that this is not an intrinsic bias in the sample selection since our sample is selected based only on the afterglow properties. Regardless, from Fig. \ref{fig:Sample_properties}, we conclude that our sample primarily traces the low-mass end of the galaxy mass function at the respective redshifts (by comparing against the characteristic stellar mass in the Schechter function). \\
\vspace{-2ex}

\noindent 
2. While there is a significant spread, the majority of the GRB hosts in our sample are within 0.5 dex (i.e. 3$\times$) of the star formation main sequence at their respective redshifts (within observational uncertainties). It should also be noted that 
majority of the GRB hosts here are below the main sequence. 
Thus, our sample traces a moderately sub-main sequence galaxy population at $z \sim 2-6$.

\subsection{Blue-wing column density and outflows}\label{sec:blue_wing_column}

To quantify outflows, we use the multi-component Voigt-profile fits to the high-resolution GRB afterglow absorption spectra (in the rest-frame UV) and the resulting column densities from \cite{gatkine2019cgm}. We then integrate the apparent column density (derived from the fit) bluewards of $-100$ km s$^{-1}$. We define this quantity as blue wing column density ($\mathrm{N_{out}}$), which is a measure of the galactic outflow. This velocity threshold is carefully chosen to minimize any contamination from the line-of-sight absorption in the interstellar medium. A detailed justification for this limit is provided in \cite{gatkine2019cgm} through kinematic and geometric modeling of the ISM $+$ CGM of a representative galaxy in this sample (see sections 3.1, 5, 7.4,  and Appendix B in \citealt{gatkine2019cgm}). This is similar to down-the-barrel observations of outflows, albeit with random sightlines and  using high-resolution and high signal-to-noise spectra.

 We compare the blue wing column density as described above with host galaxy properties ($\mathrm{M_{*}}$ and SFR). In particular, we focus on four species. These include two high ionization potential species (high-ion) $-$ C IV and Si IV and two low ionization potential species (low-ion) $-$ Si II and Fe II. Primarily, we used C IV 1550, Si IV 1402, Si II 1526, and Fe II 1608 absorption lines to trace the outflow-galaxy relations (summarized in Figs. \ref{fig:N_vs_SFR}$-$\ref{fig:NV2_vs_SF}).  These species are selected for three reasons. 1) Their absorption lines fall within the passbands over a large redshift range at $z > 2$. 2) These lines are not too weak (leading to underestimates) or not too strong (saturated). In most cases, we do not have saturation in the blue wings. 3) They allow us to compare the differences between the relations of high-ion and low-ion species with host galaxy properties. 
 
In particular, for low-ion lines, other alternatives have been used in the literature including Si II 1260, O I 1302, and C II 1334. However, we did not use them as the primary focus of the correlation investigation to avoid potential blending issues. Notably, Si II 1260 has the most severe blending issue due to S II 1259 that is essentially at a velocity offset of -200 km/s. This can be seen in Fig. \ref{fig:Stacking_comparison} showing the stacks of the respective lines in the CGM-GRB sample. Therefore, it is difficult to reliably integrate Si II 1260 for measuring the outflows. For C II 1334 and O I 1302, the blending issue (due to C II$^*$ 1335 and Si II 1304, respectively)  is less severe for the outflows. Hence we used Si II 1526 and Fe II 1608 which are free from such blending issues for the primary investigation of the correlations. We further conducted a secondary investigation with O I 1302 and C II 1334 as a consistency check. Those results are summarized in Appendix \ref{app:low-ions}.

 \subsection{Inferring correlations and hypothesis testing}

To investigate the presence of correlations between outflow and galaxy properties, we primarily focus on the parameter space of logarithms of $\mathrm{M_{*}}$, SFR, outflow column density, and maximum outflow velocity. First, we perform Kendall-$\tau$ test by using a null hypothesis that 
there is no intrinsic correlation between the two parameters.
The $1 - p$-value from Kendall-$\tau$ test gives us the confidence level at which the null hypothesis is rejected (i.e. a smaller p-value implies higher probability of the existence of a correlation). Second, we perform a linear regression to infer the best-fit line for each investigated correlation using Schmitt's binned regression \citep{schmitt1985statistical}. Note that we include all the upper (and lower) limits in both of these analyses using the astronomy survival analysis code called ASURV \citep{feigelson1985statistical, isobe1986statistical, isobe1990asurv}. The resulting best-fit and Kendall-$\tau$ p-value are shown in the correlation figures.  

Due to multiple upper limits in the stellar mass and/or star formation rates in the sample, simply using linear regression does not provide complete information on the underlying correlations and/or their spread. Therefore, we also divide the sample in two equal parts (around the median) based on the galaxy property under consideration ($\mathrm{M_*}$ or SFR) and investigate whether the sample distribution of the outflow property (eg: $\mathrm{N_{out}}$) in the two bins is consistent with being drawn from the same population. Therefore, for this hypothesis testing, our null hypothesis is that there is no correlation between the galaxy properties and outflow properties. If the null hypothesis is true, the two samples of outflow properties (eg: column density) split based on galaxy property (eg: $\mathrm{M_*}$ or SFR) are consistent with being drawn from the same population, which would imply an absence of correlation between the given outflow property and the galaxy property.

We plot the cumulative densities of both the samples, which further visually shows the distinction or similarity between the two sample distributions. To accommodate the upper limits in our observations, we calculate the cumulative distribution function (CDF) and its spread is evaluated using a survival analysis method called Kaplan-Meier estimator (\cite{isobe1986statistical, feigelson1985statistical}). A python package called {\fontfamily{qcr}\selectfont lifelines} \citep{davidson2019lifelines} is used to calculate the CDF and its 1-$\sigma$ spread using the Kaplan-Meier method. For two-sample hypothesis testing, we use the log-rank test in {\fontfamily{qcr}\selectfont lifelines}.  The resulting $p$-values describe the probability of the two samples being drawn from the same distribution (and hence, no correlation). With a stronger intrinsic correlation, the p-value is expected to be smaller. In figures \ref{fig:N_vs_SFR}$-$\ref{fig:NV2_vs_SF}, the split in samples is shown with a vertical dotted line. The median of the sample on either side and its 68-percentile spread are shown in large square points for comparison (on X-axis, it is 68-percentile spread in the sample, on Y-axis, it is 68-percentile spread in the inferred median).   

\section{Outflow correlations}
\label{sec:outflow_correlations}
In this section, we describe various outflow-galaxy correlations. A wide variety of correlations were investigated. In this section, we only describe the ones where at least one of the species show a $\gtrsim 2\sigma$ correlation. The rest of the correlations are described in Appendix \ref{app:Additional_correlations} for the sake of completeness and a visual comparison. All the investigated correlations are summarized in Table \ref{tab:CGM_GRB_correlations}.

\subsection{Outflow column density $vs$ galaxy properties}
In Figures \ref{fig:N_vs_SFR} and \ref{fig:N_vs_M*}, we plot the blue wing column densities of C IV, Si IV, Fe II, and Si II with SFR and $\mathrm{M_{*}}$, respectively. The key findings are summarized in the following subsections.

\subsubsection{Blue-wing detection fraction}
We define the detection fraction as the number of objects with detected blue-wing absorption divided by the total number of objects in each sample. The detection fractions in the left and right samples in Fig. \ref{fig:N_vs_SFR} ($\mathrm{N_{out}}$ $vs$ SFR)  are: C IV (1 and 0.92), Si IV (0.92 and 1), Fe II (0.62 and 0.71), and Si II (0.66 and 0.77). The relative difference in the left and right samples is insignificant (i.e. contributed by an excess of one non-detection in one of the samples). We find the same result Fig. \ref{fig:N_vs_M*} ($\mathrm{N_{out}}$ $vs$ $\mathrm{M_{*}}$). Thus, we conclude that the detection fractions in the outflow do not strongly depend on the galaxy's stellar mass or star formation rate. However, we note that the blue-wing detection fraction is significantly higher in high-ion species (C IV, Si IV) compared to the low-ion species, hinting at a prevalence of outflows primarily traced by the warm phase ($\mathrm{10^{4.5} - 10^{5.5}}$ K, \cite{tumlinson2017circumgalactic, gatkine2019cgm}) assuming collisional ionization equilibrium. 

\subsubsection{$N_{out}$ $vs$ SFR}
\label{subsubsection:N_vs_SFR}
\begin{figure*}
\centering
\includegraphics[width=0.95\textwidth]{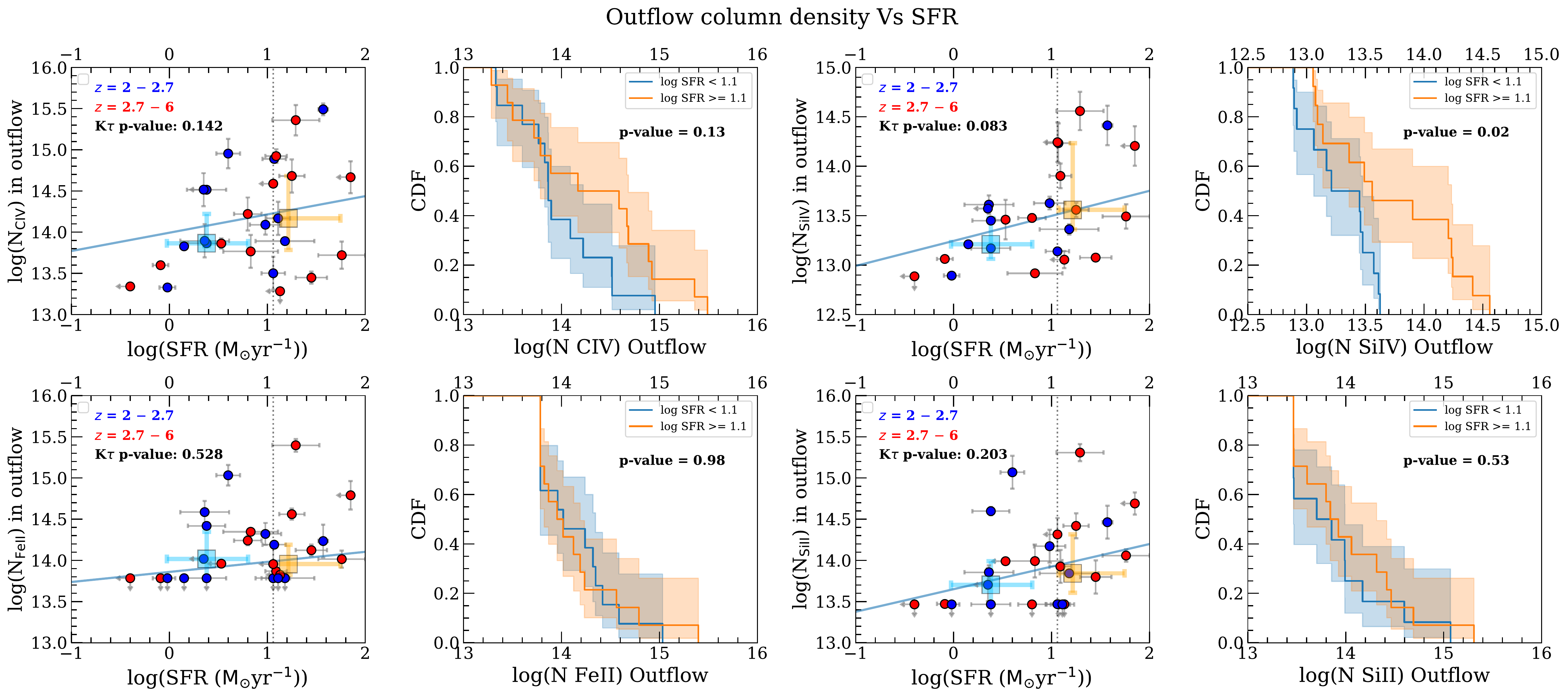}
\figcaption{\label{fig:N_vs_SFR} 
Column density in the outflows in GRB hosts traced by high-ion (C IV, Si IV) and low-ion (Fe II, Si II) species $vs$ their SFR. The Kendall-$\tau$ $p$-value indicates the strength of correlation ($1-p$ is the confidence level of the correlation). The vertical dotted line splits the sample into two equal groups around the median SFR. The CDF of each group is shown on the right to compare the distributions of low-SFR and high-SFR parts of the sample. The log-rank-test $p$-value shown in the CDF plot measures the extent to which the distributions are similar and hence consistent with no correlation. The median and 68-percentile spread of the median column density is shown using the blue and orange squares. The best-fit line (including limits in the data) is also shown here. Apart from a weak correlation, there is a significant increase in the spread of column density at high SFR (particularly for the high-ion lines). }
\end{figure*}

From visual inspection of $\mathrm{N_{out}}$ $vs$ SFR panels in Fig. \ref{fig:N_vs_SFR}, we note that there is a greater prevalence of high $\mathrm{N_{out}}$ in the high-SFR sample compared to the low-SFR sample. The CDF plots provide a quantitative measure of any such relation. First, we focus on the high-ion species (C IV and Si IV). The low-SFR and high-SFR samples are most distinct (i.e. small $p$-value) in high-ion species. We can reject the hypothesis of absence of correlation (between $\mathrm{N_{out}}$ $vs$ SFR) for C IV and Si IV with 87\% and 98\% confidence ($1-p$) respectively. Thus, a higher star formation rate is likely to be correlated with a higher column density of C IV and Si IV in the outflows (albeit with a relatively lower confidence). In addition, we note that the spread of $\mathrm{N_{out}}$ (as evident from the 68-percentile errorbars on the median points) is considerably higher in the high-SFR sample compared to the low-SFR sample (by $\sim$ 0.3-0.5 dex). 
This effect is discussed in more detail in Section \ref{subsec:Outflow_geometry}. 

Unlike the case of high-ion species, the $p$-value is high, indicating a weak (for Si II) or no correlation (for Fe II) of low-ion $\mathrm{N_{out}}$ with SFR.  The fact that we see a stronger high-ion outflow with star formation but only a weak change in the low-ion species indicates that a higher SFR leads to a stronger high-ion traced outflow but does not significantly affect the low-ion traced outflow phase. In other words, a higher SFR selectively enriches the outflow with high-ions.


    

\subsection{Outflow kinematics and galaxy properties}
We study the relationship between outflow kinematics and galaxy properties using the maximum velocity of the outflow (i.e. maximum velocity in the blue wing), $\mathrm{V_{max}}$. We define $\mathrm{V_{max}}$ as the velocity of the most blue-shifted absorption component $+$ the half-power width of that component. The maximal velocity is a key determinant of the outflow energy and mass outflow rate and hence, the enrichment of the CGM (and IGM). Therefore, in this paper, we use  $\mathrm{V_{max}}$ as a proxy for outflow kinematics to investigate the effect of galaxy properties.  

\subsubsection{Outflow $\mathrm{V_{max}}$ $vs$ SFR}\label{Subsec:Vmax_vs_SFR}
\begin{figure*}
\centering
\includegraphics[width=0.95\textwidth]{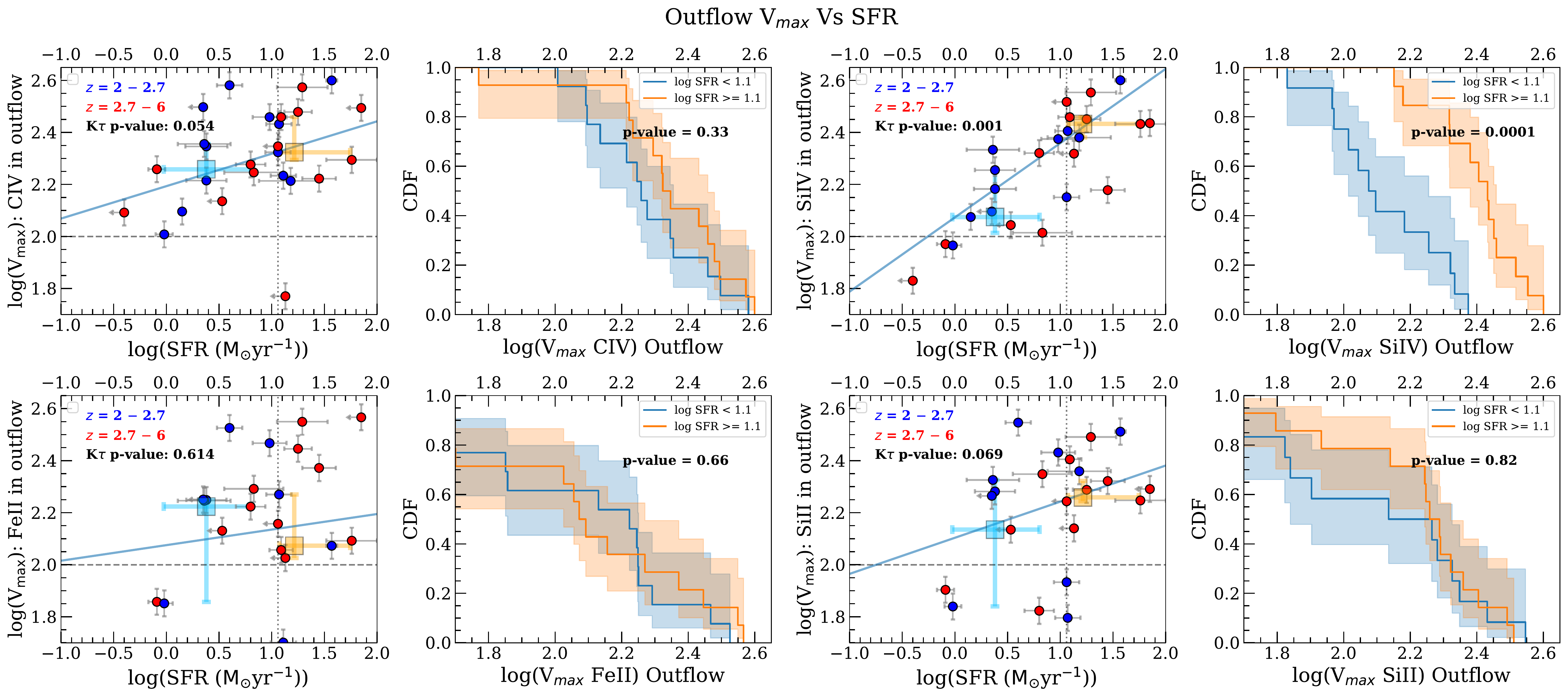}
\figcaption{\label{fig:Vmax_vs_SFR} 
Same as Figure \ref{fig:N_vs_SFR}, for the maximum outflow velocity, $\mathrm{V_{max}}$ $vs$ $\mathrm{SFR}$. The horizontal dashed line in the panels shows the 100 km s$^{-1}$ level, which we treat as the threshold for outflow.}
\end{figure*}

From Fig. \ref{fig:Vmax_vs_SFR}, a strong correlation is observed between $\mathrm{V_{max}}$ and SFR for high-ion species (2-$\sigma$ for C IV and 3-$\sigma$ for Si IV). The best-fit relations are given by $\mathrm{V_{max}}$ $\propto$ $\mathrm{SFR^{0.12}}$ and $\mathrm{SFR^{0.29}}$ for C IV and Si IV, respectively. 
This correlation is much tighter than the SFR $-$ column density relation. The smaller variation would mean the velocity gain due to higher SFR is mostly independent of the sightline being probed. By combining this with previous results from Section \ref{subsubsection:N_vs_SFR}, it can be said that star formation uniformly drives up the high-ion outflow velocity, but also imparts a large variance in the overall amount of outflowing material (column density) that is being driven. We discuss this aspect in more detail in Section \ref{subsec:Outflow_geometry}.

On the other hand, for low-ion species, the correlation is weaker (39\% and 93\% confidence for Fe II and Si II respectively), primarily due to a larger spread in the $\mathrm{V_{max}}$ compared to high-ion species (which can be easily seen by comparing their CDF plots). This shows a larger variance in the kinematics of low-ion traced outflows relative to the high-ion traced outflows. 

In addition, Fig. \ref{fig:Vmax_vs_SFR} also shows that the high-redshift and the low-redshift populations (groups $z1$ and $z2$) follow the same trend for $\mathrm{V_{max}-SFR}$ correlation. We do not observe any significant evolution in the relation of outflow kinematics and SFR. This further corroborates a previous result from \cite{gatkine2019cgm} which shows no evolution in the CGM kinematics in the same two redshift bins.

\subsubsection{Outflow kinematics $vs$ Halo mass}
\label{subsubsec:Vnorm_vs_Mhalo}
The halo mass is an important factor in determining whether the outflow will eventually escape and enrich the intergalactic medium or it will virialize and enrich the CGM. Therefore, it is important to understand how the outflow velocity compares with the characteristic velocity of the halo. To study how the outflow kinematics relate to the halo mass, we define a normalized velocity, $\mathrm{V_{norm}}$ = $\mathrm{V_{max}}$/$\mathrm{V_{circ, halo}}$, where $\mathrm{V_{circ, halo}}$ is the halo circular velocity. The $\mathrm{V_{circ, halo}}$ is calculated using the following equations from \cite{mo2002abundance}:

\begin{equation}
    V_{circ, halo} =  \left ( \frac{GM_{halo}}{r_{halo}} \right ) ^{1/2}
\end{equation}

\begin{equation}
    r_{halo} =  \left ( \frac{GM_{halo}}{100\Omega_{m}H_{0}^{2}} \right ) ^{1/3}(1+z)^{-1}
\end{equation}

Here, $\mathrm{M_{halo}}$ is calculated using the redshift-dependent stellar-to-halo mass ratio from \cite{wechsler2018connection}.  

\begin{figure*}
\centering
\includegraphics[width=0.95\textwidth]{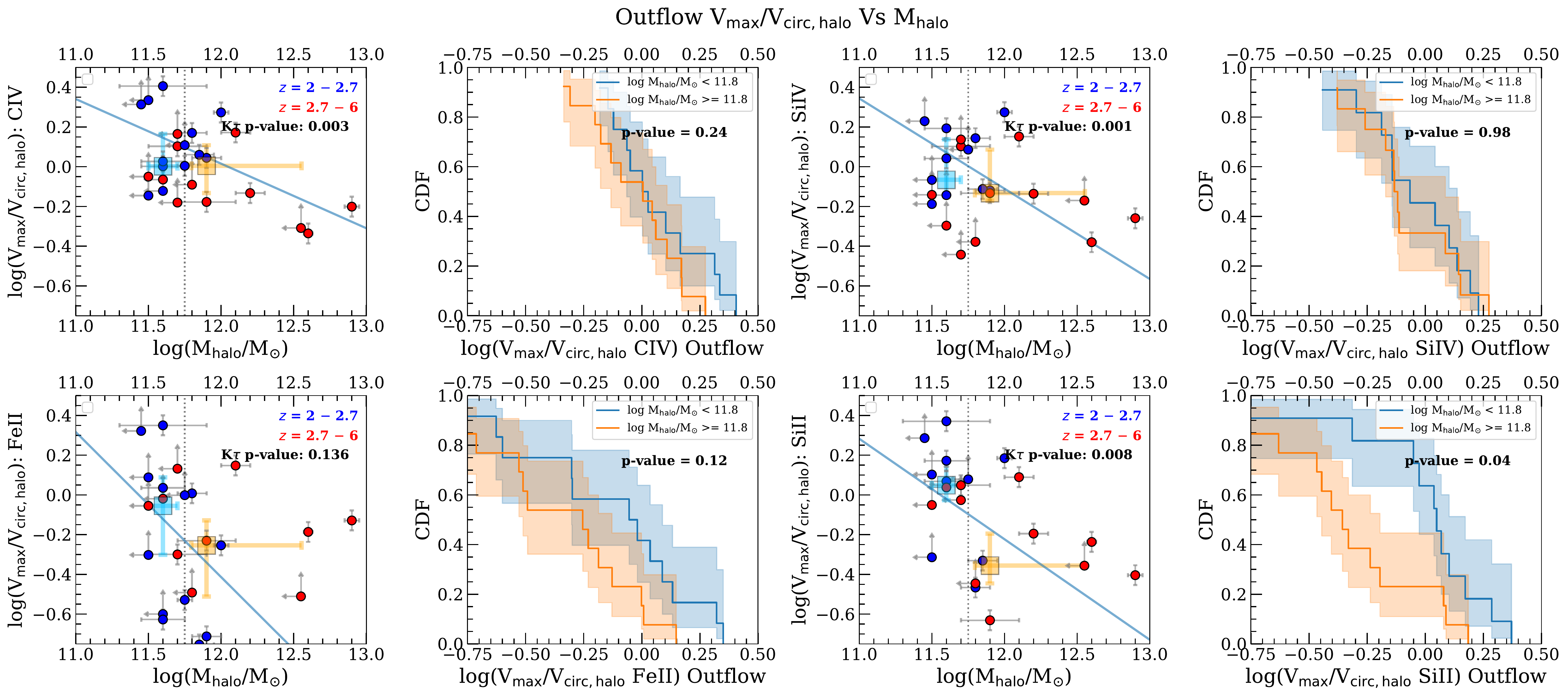}
\figcaption{\label{fig:Vnorm_vs_M_halo} 
Same as Figure \ref{fig:N_vs_SFR}, for the normalized velocity, $\mathrm{V_{max}/V_{circ, halo}}$ $vs$ $\mathrm{M_{halo}}$.}
\end{figure*}

The $\mathrm{V_{norm}}-\mathrm{M_{halo}}$ relation is summarized in Fig. \ref{fig:Vnorm_vs_M_halo}. We observe a clear inverse correlation in both high-ion and low-ion outflows. The inverse correlation is slighlty stronger in high-ions (confidence: 99.7\% in C IV, 99.9\% in Si IV) compared to low-ions (confidence: 86.4\% in Fe II, 99.2\% in Si II).
We note that most of the low-redshift points appear in the low-$\mathrm{M_{halo}}$ group while the high-redshift points appear in the high-$\mathrm{M_{halo}}$ group. This is because the stellar-to-halo mass ratio is larger at higher redshifts. Also,  $\mathrm{V_{circ}}$  scales as $\mathrm{M_{halo}^{1/3}}$, thus lowering the value of $\mathrm{V_{norm}}$ for the high-redshift objects. 

The key takeaway from the $\mathrm{V_{norm}}-\mathrm{M_{halo}}$ relation is that the outflows in low-mass halos have a greater probability of reaching and/or escaping the outer CGM and enriching the intergalactic medium. Assuming that $\mathrm{V_{max}}$ reflects the gas motion at the largest radii of the outflows, as interpreted in \cite{martin2009physical} (with or without acceleration at larger radii), we can infer that outflows with $\mathrm{V_{max}}$ $>$ $2 \times \mathrm{V_{circ}}$ (i.e. $\mathrm{log(V_{norm})} > 0.3$) are most likely to escape the CGM and enrich the intergalactic medium at high redshifts. 


\subsection{Outflow $\mathrm{V_{norm}}$ $vs$ specific SFR}

\begin{figure*}
\centering
\includegraphics[width=0.95\textwidth]{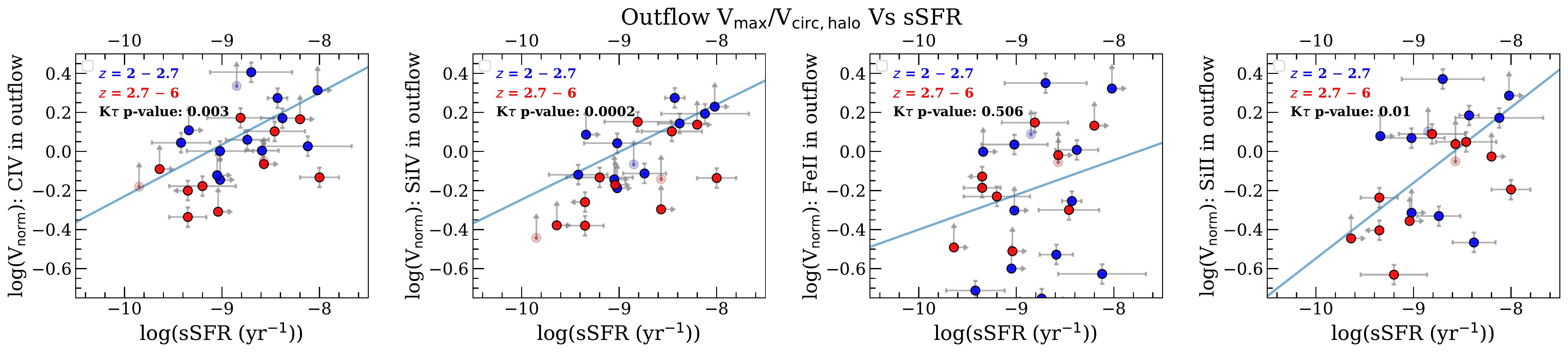}
\figcaption{\label{fig:Vnorm_vs_sSFR} 
Same as Figure \ref{fig:N_vs_sSFR}, for the the scaling relations of normalized maximum velocity ($\mathrm{V_{max}/V_{circ, halo}}$) with specific SFR (= SFR / $\mathrm{M_*}$).}
\end{figure*}

Following the strong $\mathrm{V_{norm}-M_{halo}}$ relation observed in Section \ref{subsubsec:Vnorm_vs_Mhalo}, we set out to explore whether $\mathrm{V_{norm}}$ (which is a gauge of whether the outflow can escape) is impacted by the sSFR. These results are summarized in Fig. \ref{fig:Vnorm_vs_sSFR}. We clearly observe a strong correlation between $\mathrm{V_{norm}}$ and sSFR for both high- and low-ions (except Fe II). 

Our results are analogous to those found in  \cite{heckman2016implications} for extreme starbursts at $z \sim 0-0.7$ using Si II line. They find that $\mathrm{V_{norm}}$ $\propto$ $\mathrm{sSFR^{0.25}}$ with a $>$ 3-$\sigma$ confidence. Our scaling relations are consistent with this slope ($\mathrm{V_{norm}}$ $\propto$ sSFR$^{0.26}$, sSFR$^{0.24}$, sSFR$^{0.26}$, and sSFR$^{0.39}$ for C IV, Si IV, and Si II, with 3-$\sigma$, 3.7-$\sigma$, and 2.6-$\sigma$ confidence, respectively).

We note that $\mathrm{V_{norm}}$ is larger in \cite{heckman2016implications}. This is because they study extreme starburst galaxies with typical SFRs that are higher by at least an order of magnitude compared to our sample. This leads to a $2-3$x boost in outflow velocities. Regardless, the correlation is fairly robust in the sSFR range : $\mathrm{log(sSFR)} \sim -10$ to $-7.5$, similar to our range of interest. Given the strong agreement between the slopes obtained from the low-redshift results \citep{heckman2016implications} and our high-redshift results, we argue that the $\mathrm{V_{norm}-sSFR}$ proportionality may be a redshift-independent fundamental property of star-forming galaxies. This should be investigated further using low- and intermediate-redshift analogs of high-$z$ star-forming galaxies. 

This is the first such evidence of $\mathrm{V_{norm}-sSFR}$ correlation at $z > 2$. The combined $\mathrm{V_{norm}-sSFR}$ and $\mathrm{V_{norm}}$$-$$\mathrm{M_{halo}}$ relations (Section \ref{subsubsec:Vnorm_vs_Mhalo}) imply that the outflows from the low-mass halos and high-sSFR galaxies have the highest probability of escaping the halo and transferring matter to the intergalactic medium, and thus enriching it with metals.

\subsection{Outflow rate and mass loading correlations}

Assuming a spherical outflow geometry, the time-averaged mass outflow rate ($\dot{M}_{out}$)  across a cross-section area $A$ (over a dynamical timescale $\sim$ $\mathrm{R/V_{mean}}$) can be written as: 
\begin{equation}\label{eqn:M_out_basic} 
    \dot{M}_{out} = m \frac{A}{R}\int {N_{a}(v_{out})v_{out}dv}
\end{equation}

where $m$ is the mean atomic mass per H atom,  $R$ is the radius at which the cross-section is evaluated, and $N_{a}(v)$ is the apparent column density (per unit velocity) at a velocity $v$. 

Here, we use the observable parameter $\int {N_{a}(v_{out})v_{out}dv}$ (abbreviated as N$\cdot$V) as a proxy for the mass outflow rate since the radial profile of the outflow is unknown. We use the best-fit Voigt profiles for the spectra in our sample as derived in \cite{gatkine2019cgm} and evaluate $N_{a}(v_{out})$ using the apparent optical depth method ( \citealt{savage1991analysis}, also see section 4.2 in \citealt{gatkine2019cgm}). The integration is performed leftwards of $v$ = -100 $\mathrm{km ~s^{-1}}$ to consider only the outflows. We note that using the N$\cdot$V of various high- and low-ion species as the proxy for $\mathrm{\dot{M}_{out}}$ has certain limitations. The ionization fraction of the species, the metallicity of the outflow, and the outflow geometry can vary within the redshift range and as a function of galaxy properties. Despite these variations, the N$\cdot$V provides a useful insight into the comparative outflow dynamics of the high-ion and low-ion traced outflows. Keeping these strengths and limitations in mind,  we investigate the following correlations to trace the outflow dynamics.

\subsubsection{Outflow rate 
vs SFR}\label{subsubsec:Outflow_rate_vs_SFR}

\begin{figure*}
\centering
\includegraphics[width=0.95\textwidth]{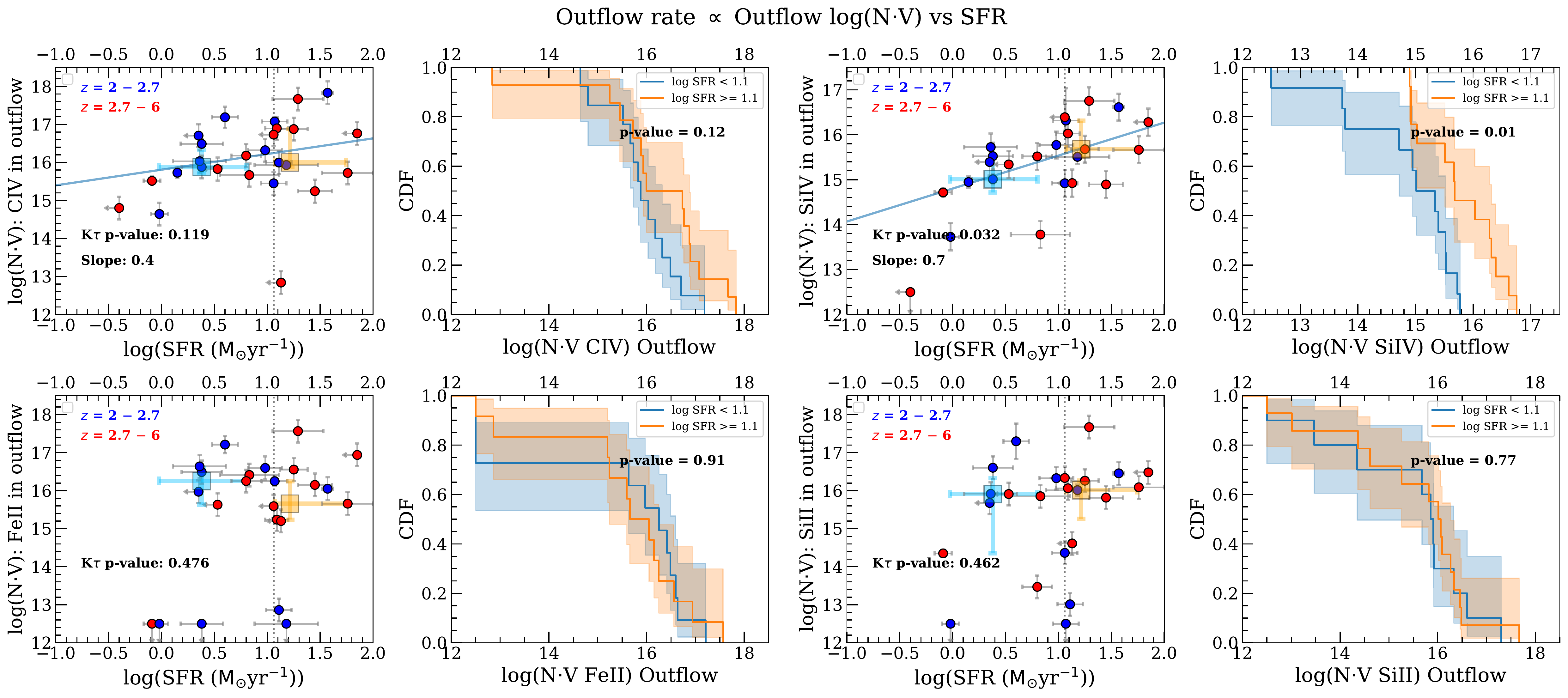}
\figcaption{\label{fig:NV_vs_SF} 
Same as Figure \ref{fig:N_vs_SFR}, for the relation of a proxy of outflow rate (N$\cdot$V = $\int N_{a}vdv$) vs $\mathrm{SFR}$. The units for Y-axis are: $\mathrm{cm^{-2}km~ s^{-1}}$. The regression fit and slope are shown where the Kendall-$\tau$ p-value is less than 0.15 (i.e. 1.5-$\sigma$ or higher level for correlation).}
\end{figure*}

In Fig. \ref{fig:NV_vs_SF}, we plot N$\cdot$V (a proxy for the mass outflow rate) against SFR. We observe an increasing trend in the high-ions (C IV and Si IV) with an approximate log-log slope of $\sim 0.5$, suggesting that the mass outflow rate of high-ion traced outflows is driven by star formation. However, we do not observe such a trend in the low-ions. This is expected given the large scatter (or absence) of $\mathrm{N_{out}}$ and $\mathrm{V_{max}}$ correlations with SFR in the low-ion species.\\ \vspace{-2ex} 

The high-ion correlation has a slope (0.4$-$0.7, see Fig. \ref{fig:NV_vs_SF}) slightly shallower than the slope obtained by \cite{heckman2015systematic} ($\mathrm{log(\dot{M}_{out})}-$log(SFR) slope of $\sim$ 1) for local starbursts using both high- and low-ion lines, albeit with a larger scatter (see Fig. 5 in their paper). \cite{chisholm2015scaling} also find a similar correlation (slope $\sim$ 0.5) in a sample of local star-forming galaxies using low-ion lines. These similarities indicate a common star-formation-driven outflow mechanism from $z$ $\sim$ 0 to $z$ $\sim$ 6. 
We note here, however, that the marked difference that we see between the high- and low-ion correlations imply a systematic difference in the outflow mechanisms of the phases traced by them. 
The blue wings in the high-ion absorption appear to be dominated by outflows driven by star formation (as supported by the correlations) while the blue wings in the low-ions are more complex. 
 
 
To gauge how the mass outflow rate compares with the star formation of the sample, we take C IV species as an example and estimate the total outflow rate as follows: 
\vspace{-2ex}
\begin{equation} \label{eqn:M_out_total}
    \dot{M}_{out, H} =  \dot{M}_{out, C IV} \frac{1}{f_{C IV} Z_{out}} \left( \frac{N_{H}}{N_{C}} \right)_{solar} 
\end{equation}

Here, $f_{CIV}$ is the ionization fraction of C IV, $Z_{out}$ is an estimate of the outflow metallicity, and $N_{C}/N_{H}$ is the carbon abundance \citep{lodders2003solar}. Using equations \ref{eqn:M_out_basic} and \ref{eqn:M_out_total},
\begin{equation}\label{eqn:mass_outflow_rate}
    \begin{aligned} 
    \dot{M}_{out} \approx  1  ~ \mathrm{M_{\odot}~yr^{-1}}  \left( \frac{(N.V)_{CIV}}{10^{16.5} cm^{-2}km/s} \right) \left (\frac{5~\mathrm{kpc}}{R_{out}}  \right )  \\
    \left (\frac{0.3}{f_{C IV}}  \right ) 
     \times \left (\frac{0.25}{Z_{outflow}}  \right ) \left (\frac{2.8 \times 10^{-4}}{(N_{C}/N_{H})_{\odot}}  \right )
    \end{aligned}
\end{equation}

Here, a representative N$\cdot$V value is taken for C IV species at SFR $\sim$ 10 $\mathrm{M_{\odot}yr^{-1}}$ (see Fig. \ref{fig:NV_vs_SF}). There is a scatter of about an order of magnitude in either direction in the observed N$\cdot$V values for C IV. The values of $Z_{outflow}$ and $f_{CIV}$ are assumed to be the same as those in \cite{gatkine2019cgm} and the radius of the outflowing shell is assumed to be 5 kpc. Note that this mass outflow rate is a lower limit given the conservative assumptions of C IV ionization fraction (see Sec 6.1.2 in \citealt{gatkine2019cgm} for a detailed explanation) and outflow radius. The measured outflow properties can arise from a radius as large as 50 kpc (i.e. the typical virial radius of these galaxies). Thus, the $\dot{M}_{out}$ estimated in Eqn \ref{eqn:mass_outflow_rate} can be larger by almost an order of magnitude. Regardless, given the scatter in the observed N$\cdot$V, many of the observed systems have an estimated mass outflow rate  comparable or even greater than the SFR.    

Given that N$\cdot$V $\propto$ $\mathrm{SFR^{0.4}}$ for C IV (or, $\mathrm{SFR^{0.7}}$ for Si IV), it is evident that the lower-SFR galaxies in our sample experience a higher mass outflow rate as a fraction of their SFR and thus undergoing an efficient removal of the gas compared to high-SFR galaxies.




\subsubsection{Mass loading factor}\label{subsubsec:Mass_loading_factor}

\begin{figure*}
\centering
\includegraphics[width=0.95\textwidth]{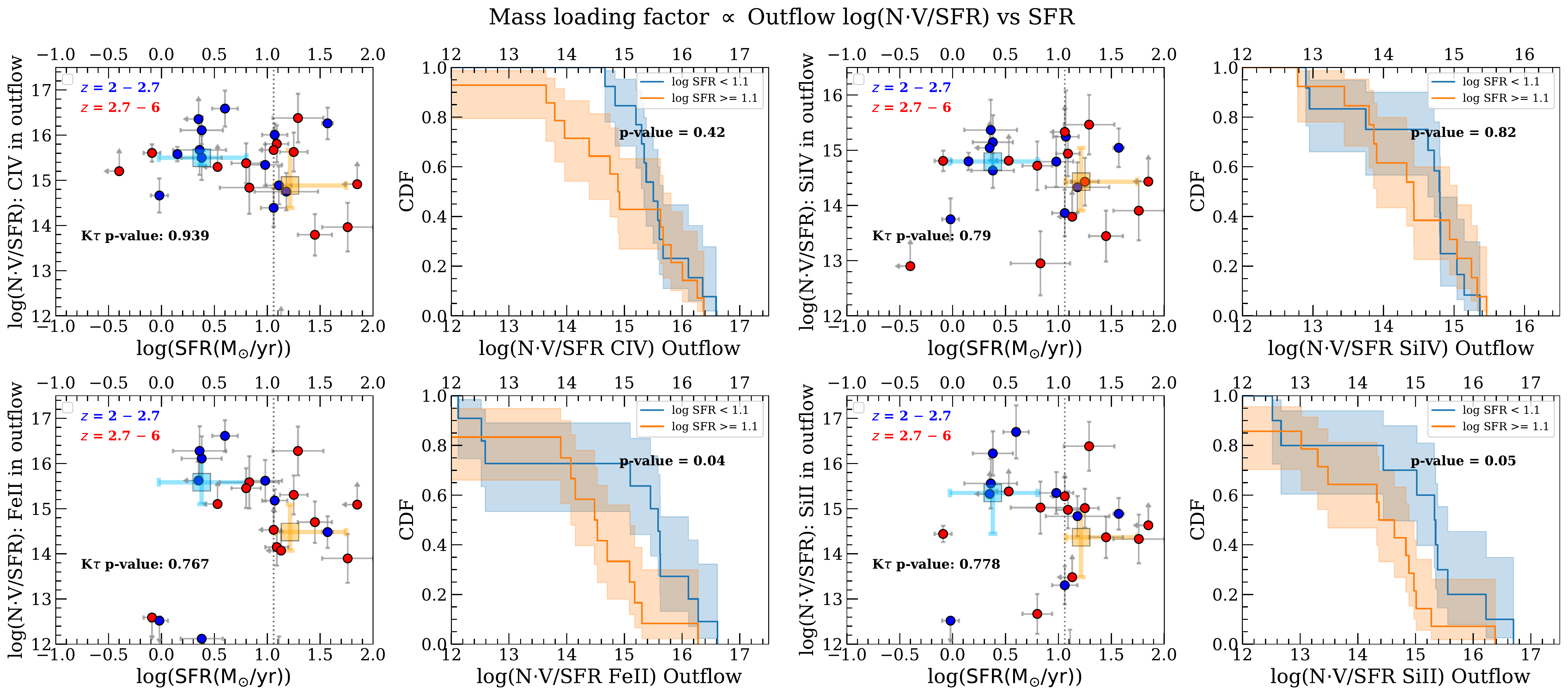}
\figcaption{\label{fig:Eta_vs_SF} 
Same as Figure \ref{fig:N_vs_SFR}, for the relation of a proxy of mass loading factor (N$\cdot$V/SFR = $\int N_{a}vdv$/SFR) vs $\mathrm{SFR}$. The units for Y-axis are: $\mathrm{cm^{-2}km~ s^{-1} M_{\odot}^{-1} yr }$. }
\end{figure*}

\begin{figure*}
\centering
\includegraphics[width=0.95\textwidth]{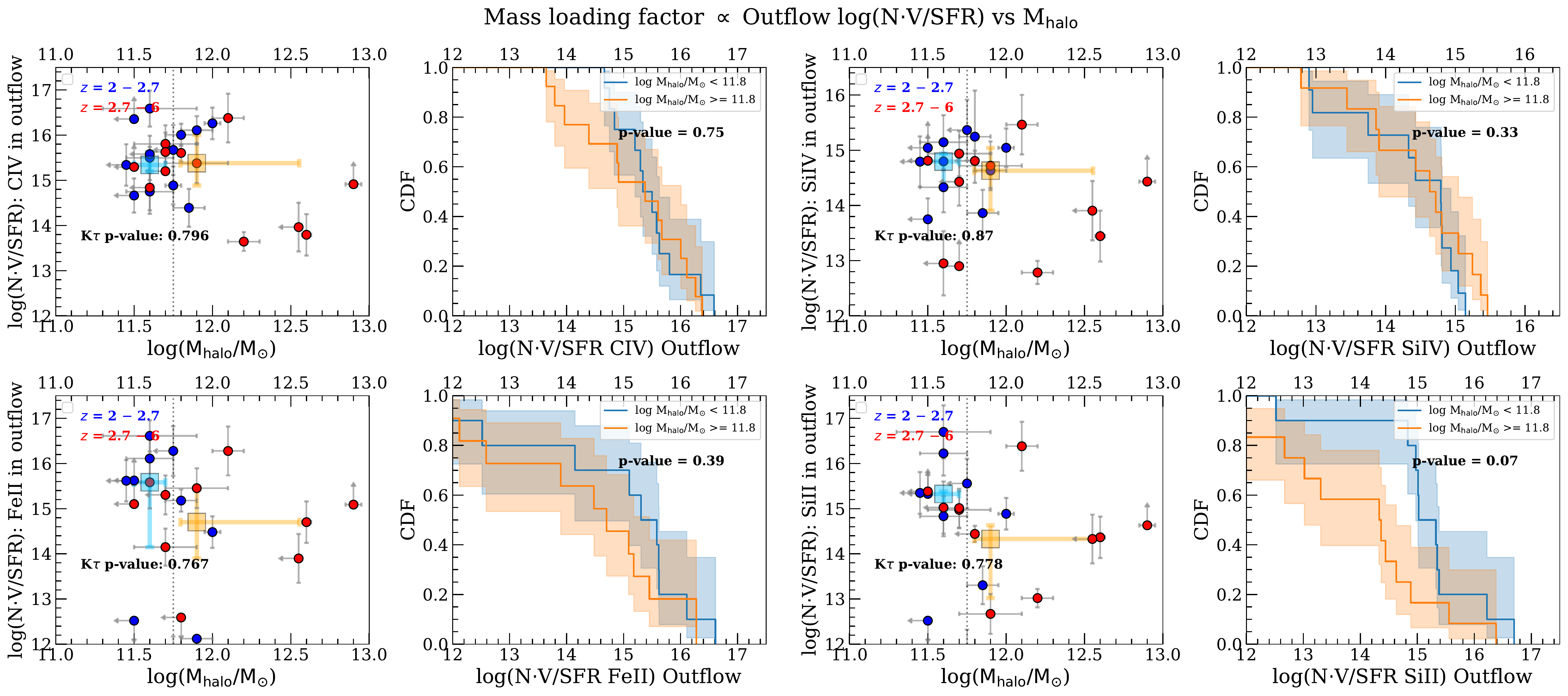}
\figcaption{\label{fig:Eta_vs_Mhalo} 
Same as Figure \ref{fig:N_vs_SFR}, for the relation of a proxy of the mass loading factor (N$\cdot$V/SFR = $\int N_{a}vdv$/SFR) vs $\mathrm{M_{halo}}$. The units for Y-axis are: $\mathrm{cm^{-2}km~ s^{-1} M_{\odot}^{-1} yr }$. }
\end{figure*}

Similar to section \ref{subsubsec:Outflow_rate_vs_SFR}, we use N$\cdot$V/SFR as the proxy for mass loading factor ($\eta$ = $\mathrm{\dot{M}_{out}/SFR}$)  to measure how efficiently the galaxy removes gas (in comparison to star formation rate). We study its relation with SFR and $\mathrm{M_{halo}}$ in Figs. \ref{fig:Eta_vs_SF} and \ref{fig:Eta_vs_Mhalo} respectively to understand the potential drivers of the mass-loading factor. 

For the $\eta~vs$ SFR plot (Fig. \ref{fig:Eta_vs_SF}), we do not find a clear correlation. However, by splitting the sample into two equal high-SFR and low-SFR bins and comparing the medians, we observe a declining trend with SFR. This suggests a  higher outflow efficiency in the low-SFR galaxies, particularly for the outflow traced by low-ion lines.  We note that this result has to be seen in conjunction with the N$\cdot$V $vs$ SFR plot (Fig. \ref{fig:NV_vs_SF}). The weaker decline in the high-ions is driven by the intrinsic correlation between N$\cdot$V and SFR. On the other hand, the stronger decline seen in the low-ions is due to a lack of such an intrinsic  N$\cdot$V$-$SFR correlation, leaving the 1/SFR as the dominant term. The observed slow decline in the high-ions is in agreement with the simulated slope for stellar feedback ($\eta$ $\sim$ SFR$^{0.35}$ for winds leaving the ISM) in the star-forming galaxies at $z~ > ~2$ in the EAGLE simulations (\citealt{mitchell2020galactic}, see Fig. 3 therein).  

In the $\eta~vs~\mathrm{M_{halo}}$ plot (Fig. \ref{fig:Eta_vs_Mhalo}), we observe a large scatter in this relation and thus, we cannot find any statistically significant correlation for either low- or high-ion species. The large scatter in this relation makes a comparison with cosmological simulations difficult. There is a possible hint for a declining outflow efficiency at high halo masses, particularly for the low-ions (as seen from the binned sample), however, a significantly larger sample is needed in the future to constrain any underlying correlation.


\subsubsection{Momentum flux vs SFR}

The stellar processes (supernovae and stellar winds) can drive momentum flux as a combination of both radiation pressure \citep{murray2005galaxies} and ram pressure of a hot outflow collectively created from the ejecta of massive stars \citep{chevalier1985wind, veilleux2005galactic, heckman2016implications}. Similar to Eqn. \ref{eqn:M_out_basic}, the time-averaged momentum flux across a cross-section area $A$ (over a dynamical timescale $\sim$ $\mathrm{R/V_{mean}}$) can be written as: 
\begin{equation}\label{eqn:p_out_basic} 
    \dot{p}_{out} = m \frac{A}{R}\int {N_{a}(v_{out})v^{2}_{out}dv}
\end{equation}

\begin{figure*}
\centering
\includegraphics[width=0.95\textwidth]{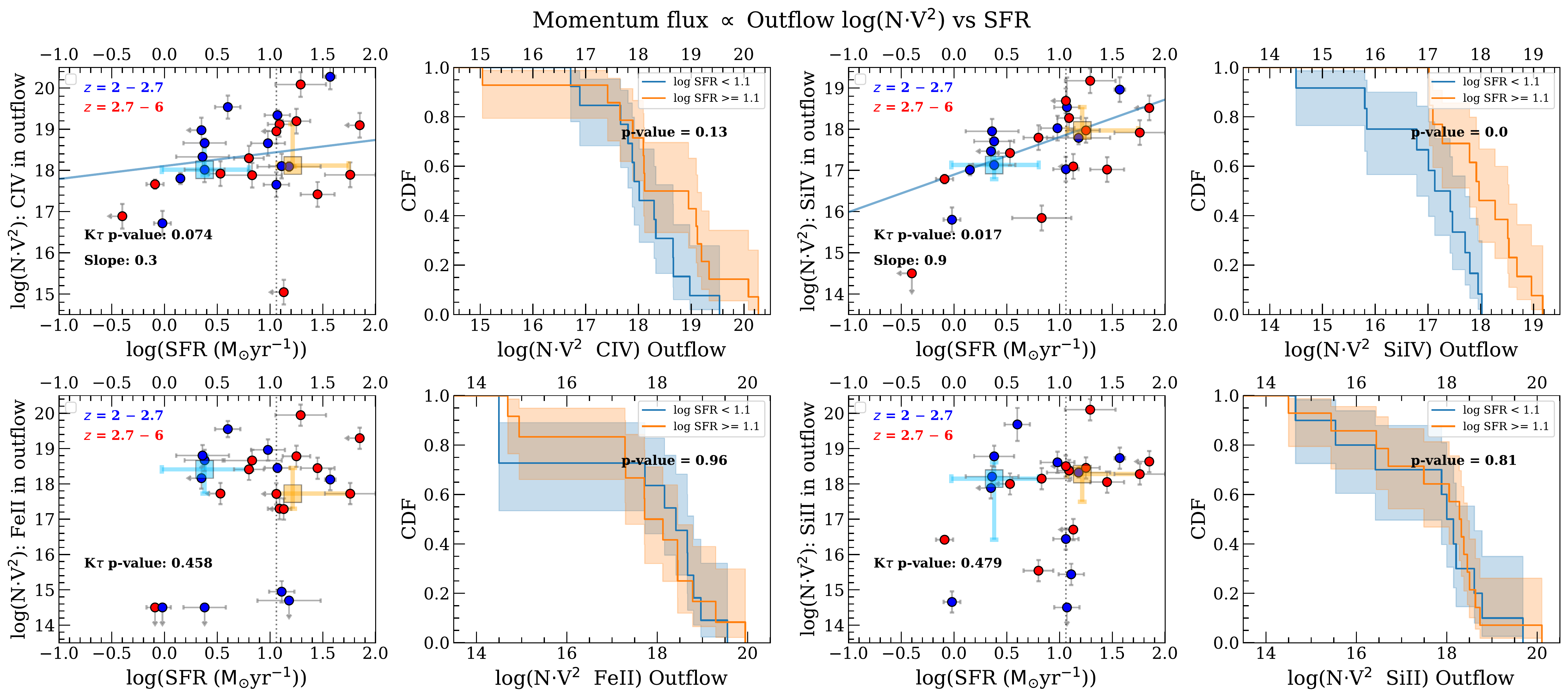}
\figcaption{\label{fig:NV2_vs_SF} 
Same as Figure \ref{fig:N_vs_SFR}, for the relation of a proxy of momentum flux (N$\cdot$V$^2$ = $\int N_{a}v^{2}dv$) vs $\mathrm{SFR}$. The units for Y-axis are: $\mathrm{cm^{-2}(km~ s^{-1})^{2}}$. The regression fit and slope are shown where the Kendall-$\tau$ p-value is less than 0.15 (i.e. 1.5-$\sigma$ or higher level for correlation).}
\end{figure*}

\begin{figure*}
\centering
\includegraphics[width=0.95\textwidth]{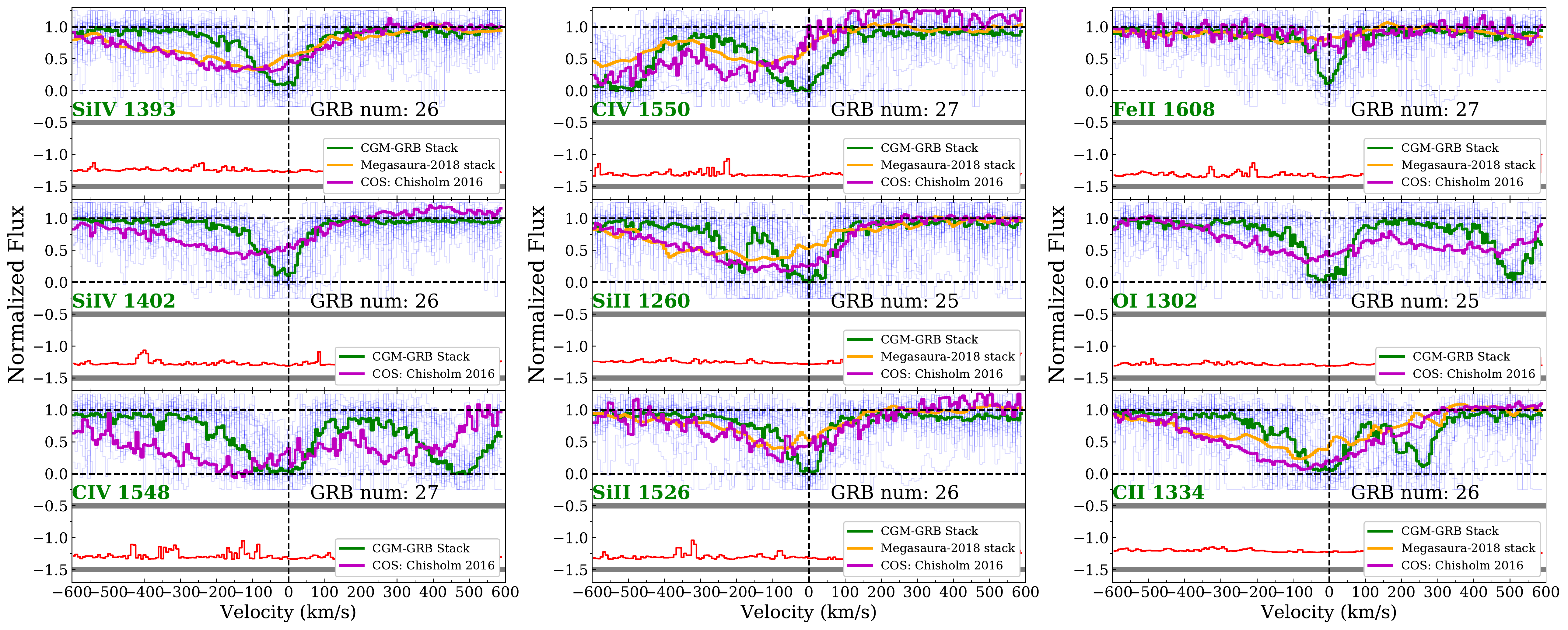}
\figcaption{\label{fig:Stacking_comparison}
The median absorption line profile for the CGM-GRB sample (in green) of the individual spectra (in blue) in the CGM-GRB sample. The red line in the bottom panel shows the rms noise (shifted vertically by $-1.5$). The overlaid traces are stacked median profiles of the MEGaSaURA $z \sim 2$ sample of 14 lensed starburst galaxies (in orange) and $z \sim$ 0.1 COS-sample of 41 starburst / star-forming  galaxies (in magenta), both sourced from \cite{rigby2018magellan}.}
\end{figure*}

We use the observable parameter $\int {N_{a}(v_{out})v^{2}_{out}dv}$ (abbreviated as N$\cdot$V$^{2}$) as a proxy for the momentum flux. In Fig. \ref{fig:NV2_vs_SF}, we plot N$\cdot$V$^{2}$  against SFR. We observe a correlation for high-ions: $\dot{p}_{out}$ $\propto$ $\mathrm{SFR^{0.3}}$ for C IV and $\mathrm{SFR^{0.9}}$ for Si IV, with a stronger correlation for Si IV.  However, we find no correlation for the low-ions. This is expected, given the strong correlation between $\mathrm{V_{max}}$ and SFR for high-ions observed in Sec. \ref{Subsec:Vmax_vs_SFR} and a lack of such a correlation for low-ions.

Next, we compare the momentum flux with star-formation-driven momentum flux to examine whether it is sufficient to drive the observed momentum flux in the high-ion traced outflows.  For a standard Kroupa/Chabrier initial mass function and a constant  star-formation rate, the estimated stellar-driven momentum flux is given by the sum of radiation pressure and ram pressure components \citep{heckman2015systematic} as:

\begin{equation}\label{eqn:momentum_flux_SFR}
\dot{p}_{*} = 4.8 \times 10^{33} ~\mathrm{dynes} \times \mathrm{SFR}~ \mathrm{(\mathrm{in ~M_{\odot} yr^{-1}})}
\end{equation}



To compare this with the observed momentum flux, we use C IV ion as a probe and rewrite Eqn. \ref{eqn:p_out_basic} as follows:  

\begin{equation}\label{eqn:momentum_flux_total}
    \begin{aligned}
    \dot{p}_{out} =  5 \times 10^{32} ~\mathrm{dynes} \left(\frac{N.V^{2}}{10^{18.5} cm^{-2}km^{2}/s^{2}} \right)   \left (\frac{0.3}{f_{C IV}}  \right ) \\
     \times \left (\frac{0.25}{Z_{outflow}}  \right ) \left (\frac{2.8 \times 10^{-4}}{(N_{C}/N_{H})_{\odot}}  \right )
    \end{aligned}
\end{equation}

Here, a representative N$\cdot$V$^{2}$ value is taken for C IV species at SFR $\sim$ 10 $\mathrm{M_{\odot}yr^{-1}}$ (see Fig. \ref{fig:NV2_vs_SF}). There is a scatter of about an order of magnitude in either direction in the observed N$\cdot$V$^2$ values for C IV. The values of $Z_{outflow}$ and $f_{CIV}$ are assumed to be the same as those in \cite{gatkine2019cgm} and the radius of the outflowing shell is assumed to be 5 kpc. We issue the same caution as in Sec. \ref{subsubsec:Outflow_rate_vs_SFR} about this estimate being a lower limit given the conservative assumptions of C IV ionization fraction and outflow radius.  

Despite the scatter in the observed N$\cdot$V$^2$, it is clear that the estimated momentum flux (from Eqn. \ref{eqn:momentum_flux_total}) is smaller than that driven by star formation (from Eqn. \ref{eqn:momentum_flux_SFR}). Thus, it is evident that the observed high-ion traced outflows are primarily driven by the momentum directly injected by supernovae and stellar winds. This is different from the results on the low-redshift starburst and star-forming galaxies (\citealt{heckman2015systematic} and \citealt{chisholm2017mass}, respectively), where the momentum flux in the outflows is comparable or even greater than that imparted by star formation for a significant fraction of the sample. However, the slope of the N$\cdot$V$^2-$SFR (or $\dot{p}_{out}-$SFR) correlation for high-ions (particularly Si IV) is consistent with the local starburst result from \cite{heckman2015systematic}. In the future, we will perform detailed ionization modeling to further examine these correlations.





\section{Discussion}
\label{sec:discussion}

\subsection{Star-formation-driven outflow}
The strong correlation of the blue-wing  column density and maximum outflow velocity with SFR, as seen in Figs. \ref{fig:N_vs_SFR} and \ref{fig:Vmax_vs_SFR} provides an important evidence for a star-formation-driven outflow. If we consider Si IV as the outflow tracer, the best-fit lines suggest $\mathrm{N_{SiIV}}$ $\propto$ SFR$^{0.25}$  and $\mathrm{V_{max, SiIV}}$ $\propto$ SFR$^{0.29}$. Our $\mathrm{V_{max}}$ slope is in close agreement with the slope derived in \cite{sugahara2017evolution} for $z \sim 2$ star-forming galaxies (= 0.25). 

Similarly, the (weak) trends of high-ion column density and $\mathrm{V_{max}}$ with $\mathrm{M_{*}}$ are consistent with previous observations of C IV-traced outflows from \cite{du2018redshift} at $z \sim 1-1.35$. However, our sample extends to 0.4 dex lower mass, where we start to see a systematic decline in $\mathrm{N_{out}}$ as well as $\mathrm{V_{max}}$ which is not seen in these previous observations. We  argue that this decline is driven by the lower star formation rates. Similarly, the large spread (in $\mathrm{N_{out}}$ and $\mathrm{V_{max}}$ ) observed at  log($\mathrm{M_*/M_{\odot}}$) $>$ 9.4 is due to the large spread in SFR at this mass range in our sample, as shown in Fig. \ref{fig:Sample_properties}. Hence, we conclude that the apparent trends between $\mathrm{V_{max}}$ or $\mathrm{N_{out}}$ and $\mathrm{M_*}$ are almost entirely modulated by the SFR.

For low-ion outflows, the best-fit relation for Si II is $\mathrm{V_{max, SiII}}$ $\propto$ SFR$^{0.14}$. Various surveys from low to high redshifts have reported correlation between some form of $\mathrm{V_{max}}$ for low-ion species and SFR. Our results agree with the slopes observed in  \cite{weiner2009ubiquitous, bradshaw2013high, bordoloi2014dependence, chisholm2015scaling} with redshifts ranging from $z \sim 0 - 1.6$. Our correlations also qualitatively agree with $z \sim 0.5$ sample of \cite{rubin2014evidence} (only for their galaxies without systemic absorption). From the $\mathrm{V_{max}-SFR}$ slope derived here, we further corroborate the suggestion from \cite{sugahara2019fast} that the $\mathrm{V_{max}-SFR}$ could be a more fundamental relation over a wide redshift range ($z \sim 0-6$) for star-forming main sequence. However, we caution that there is a significant spread in the relation depending on the species used for deriving them (for instance, we get a slope of 0.29 for Si IV and 0.14 for C IV).    

Both \cite{erb2012galactic} and \cite{rubin2012direct} find a strong correlation of $\mathrm{V_{max}}$ with stellar mass and weak correlation with SFR for galaxies of similar mass and SFR range to our sample in the redshift ranges $z \sim 1-2$ and $z \sim 0.3-0.7$ for low-ion species. Similarly, \cite{bordoloi2014dependence} and \cite{chisholm2015scaling} find a high-significance correlation with $\mathrm{M_*}$ using Mg II and Si II species respectively. \cite{rubin2012direct} argue that this could be because star formation history and/or galaxy dynamics have a more direct physical link to maximum wind velocities than current star formation activity. However, we find that the trends flip $-$ a stronger correlation with SFR compared to stellar mass (for instance, consider Si II in Figs. \ref{fig:Vmax_vs_SFR} and \ref{fig:Vmax_vs_M*}). This indicates that for low-mass galaxies at $z > 2$, the current star formation has a greater impact on the observed low-ion outflows than its star formation history. This is interesting from the perspective of causal connection. The timescale required for a 250 km s$^{-1}$ outflow to travel 50 kpc (roughly the virial radius of typical galaxies at $z \sim 3$) is about 200 Myr, while the UV-based SFR that we measure is from the past 100 Myr. This could indicate a long-lasting star formation activity. This is in line with our typical depletion timescale of 500 Myr (using ($\mathrm{M_{gas}} \sim \mathrm{M_{*}}$)/SFR $\sim$ 1/sSFR).

We compare our scaling relations with the recently published results from the TNG50 simulations \citep{nelson2019first}. We find that the slopes of our $\mathrm{V_{max}}-\mathrm{SFR}$ relation (for both high and low ions, slope $\sim$ 0.12$-$0.28) are consistent with the slope of $0.15-0.2$ in \cite{nelson2019first} (see Fig. 15 therein). However, our results do not fully agree with the predicted $\mathrm{V_{max}}-\mathrm{M_*}$ relation. While their slope of 0.2 at $z = 2$ is consistent with our high-ion results (slope = 0.19 for Si IV), \cite{nelson2019first} show a slope that steepens with redshift (eg: slope = 0.3 at $z = 4$). We do not observe such steepening in our $\mathrm{V_{max}}-\mathrm{M_*}$ plots (Fig. \ref{fig:Vmax_vs_M*}). In fact, the slope appears to be shallower in the high-redshift group (group $z2 \sim 2.7 - 6$). This discrepancy may be explained by a combination of two factors: a) the increasing contribution of hotter phases in the outflow with increasing $\mathrm{M_*}$ (see Fig. 10 in \cite{nelson2019first}) and b) the outflow scaling relations shown in \cite{nelson2019first} include all the phases while our observations might only cover the warm phases for high ions (assuming collisional ionization equilibrium).

Given the strong $\mathrm{N_{out} - SFR}$ and $\mathrm{V_{max} - SFR}$ correlations, particularly for high-ion outflows, it is expected that the mass outflow rate would be strongly correlated with SFR. In Sec. \ref{subsubsec:Outflow_rate_vs_SFR} and Fig. \ref{fig:NV_vs_SF}, we use N$\cdot$V (= $\int {N_{a}(v_{out})v_{out}dv}$) as a measure of the mass outflow rate and obtain the following relations, 

\begin{equation}
\begin{aligned}
\mathrm{\dot{M}_{out}} \propto \mathrm{SFR^{0.4}} ~\mathrm{for ~C ~IV}\\
\mathrm{\dot{M}_{out}} \propto \mathrm{SFR^{0.7}} ~\mathrm{for ~Si ~IV}
\end{aligned}
\end{equation}


The slope obtained here is in good agreement with the slope observed in the FIRE simulations ($\sim$ 0.6) at $4 > z > 2$ with the rate evaluated at 0.25$\mathrm{R_{vir}}$ (see Fig. B2 in \cite{muratov2015gusty}). This slope is also consistent with EAGLE simulations at similar redshift range ($z \sim 2.4-4.7$) for gas particles ejected out of the ISM through galactic winds (see Fig. 3 in \cite{mitchell2020galactic}). Both of these slopes have been obtained in the log(SFR) range of -0.5 to 1.5, similar to our SFR range.  


\subsection{Stacking comparison with previous studies}\label{sec:stacking_comparison}
We compare the stacked profiles of the high- and low-ion species in the CGM-GRB sample against the publicly available stacks of high- and low-redshift down-the-barrel outflow studies from \cite{rigby2018magellan}. This includes a $z \sim 2$ sample of 14 gravitationally lensed starburst galaxies (MEGaSaURA sample) and a sample of 41 starburst / star-forming (majority starburst) at $z \sim 0.1$ (the COS-sample , \citealt{chisholm2016shining}).

We observe that the outflows in  both MEGaSaURA and COS-sample stacks $-$ which consist of  starbursts (SFR $\sim$ 10$\times$ main sequence) at their respective redshifts $-$ are considerably broader and stronger compared to the CGM-GRB sample, where majority of the galaxies at $\sim$0.5 dex below the main sequence at their respective redshifts (see Fig. \ref{fig:Sample_properties}). Further, we note that the galaxies in the $z \sim 0.1$ COS-sample have a similar $\mathrm{M_*}$ and SFR to our CGM-GRB host galaxies at $z > 2$ (see \citealt{chisholm2016shining}), yet the COS-sample outflows are significantly stronger and broader than the CGM-GRB outflows for both high- and low-ions. This shows that, over a broad redshift range, the strength of the outflow is correlated with the main-sequence offset at the respective redshifts rather than the simple star-formation rate. This should be a critical consideration while comparing outflow samples at different redshifts.

On the other hand, the MEGaSaURA sample galaxies have a similar redshift range ($z \sim $ 1.68$-$3.6) to our CGM-GRB sample, but are significantly above the main sequence (and hence, starburst). Similar to the COS- sample, their outflows are significantly broader in velocity  and stronger in absorption depth for both high- and low-ions.  This comparison further corroborates  that the SFR-driven outflows are strongly correlated  with the  main-sequence offset  at the respective redshifts.

We should, however, caution that a direct stacking comparison of the line profiles with previous down-the-barrel samples in the literature is an unfair comparison due to several reasons. The most important is the selection effect. By definition, down-the-barrel samples at $z > 2$ are selected to be galaxies bright enough in rest-frame UV for absorption spectroscopy ($g_{AB}$ $<$ 21, for MEGaSaURA). On the other hand, the GRB hosts in our sample are faint, with apparent magnitude $>$ 23 in most cases. In fact, the low-mass, low-SFR galaxies at high-z are extremely difficult to probe at high resolution using down-the-barrel technique. However, GRB sightlines offer a unique opportunity to probe this population, which is not feasible at scale by any other method.

Lastly, due to a combination of selection effect and galaxy properties, the down-the-barrel observations at high-z preferentially look directly down the outflows. The GRB sightlines, on the other hand, are random. These reasons also contribute to the marked difference seen in the stacks of CGM-GRB sample and down-the-barrel high-z stacks (eg: MEGaSaURA) and should be studied in detail in the future.

\subsection{Evidence for High-ion traced outflows}
We observe three key differences between high-ion and low-ion outflows: 1) The detection fraction of low-ion outflows is lower than high-ion outflows ($\sim$ 65\% $vs$ 95\%), irrespective of the SFR. 2) The correlations of high-ion $\mathrm{N_{out}}$ and $\mathrm{V_{max}}$ with SFR are stronger compared to low-ions (see Figs. \ref{fig:N_vs_SFR} and \ref{fig:Vmax_vs_SFR}). Also, the spread in $\mathrm{V_{max}}$ is higher in low-ions. 3) Low-ion outflows show a much steeper decline in $\mathrm{V_{norm}}$ with higher halo masses compared to high-ion outflows (see Fig. \ref{fig:Vnorm_vs_M_halo}). These difference point towards a systematic difference in the high-ion and low-ion outflows. In addition, we can infer that the outflows in our sample are primarily traced by high-ions. 

It is well known that galactic outflows are multiphase in nature and the aforementioned differences can help understand the phase structure of the outflows. Cosmological simulations find a complex shift in the temperature distribution of the outflow with the dominant phase shifting towards higher temperatures as the stellar mass increases at $z \sim 2$ (see Fig. 10 in \cite{nelson2019first}). This trend is also seen in FIRE simulations with the contribution of $\mathrm{T} > 10^{5.3}$ K phase increasing and that of $10^4 < \mathrm{T} < 10^{4.7}$ K phase decreasing with an increasing halo mass at $z \sim 2$ (in the range $\mathrm{M_{halo}}$ $\sim$ $10^{11} - 10^{12}$ $\mathrm{M_{\odot}}$ (see Fig. A1 in \cite{hafen2019origins}). These effects can potentially explain the sharp decline in low-ion $\mathrm{V_{norm}}$ and their weaker and shallower correlations in terms of $\mathrm{N_{out}}$. Photo-ionization modeling of the observations would help understand whether this explanation is correct. Regardless, we infer that the correlations seen in the star-forming galaxies using random, narrow sightlines (as offered by GRBs) imply high-ion dominated outflows in the star-forming galaxies at $z > 2$. 

\subsection{ Outflow Geometry}
\label{subsec:Outflow_geometry}
The sharp increase in the spread of outflow column density at SFR $\gtrsim$ 10 $\mathrm{M_{\odot} yr^{-1}}$, particularly for high-ion lines is an indication of systematic variance in the outflow properties with SFR. 

There are various possible causes that could lead to an elevated spread in the apparent outflow column density. Some of the scenarios include a variance in the metallicity of the outflowing clouds due to inefficient mixing of metals \citep{schaye2007large}, a variance in the entrainment efficiency of the ISM, or the onset of wind-stimulated condensation in the CGM (see \cite{heckman2017cos}) at high star formation rates. A more careful treatment of the physical processes in the outflows is warranted to explain this phenomenon.

\begin{figure}
\centering
\includegraphics[width=0.45\textwidth]{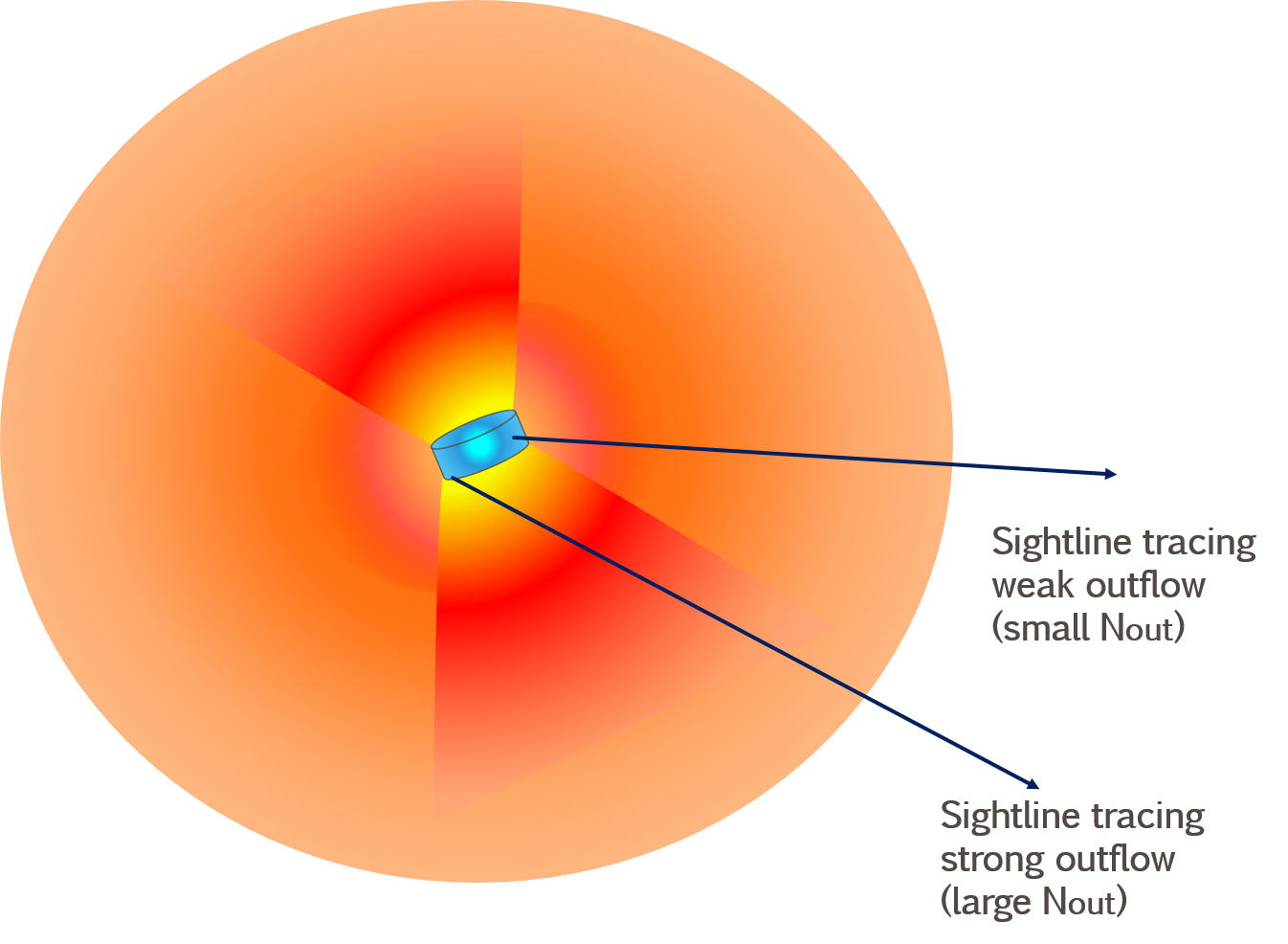}
\figcaption{\label{fig:Outflow_geometry} 
A schematic showing a stronger biconical outflow along the minor axis and weaker spherical outflow elsewhere. The narrow GRB sightlines are shown as arrows. Some of the narrow and randomly pointed GRB sightlines will  trace the strong biconical outflow, giving a larger outflow column density, while others will trace the weaker outflow, giving a smaller outflow column density. This will produce a scatter in $\mathrm{N_{out}-SFR}$ relation.}
\end{figure}

Another possible explanation for such a variance is outflow geometry. Given that GRBs sample a very narrow beam in a random direction offset from the galactic center, a spherical outflow is less likely to lead to the observed spread. In a  non-spherical outflow, with certain regions of high gas entrainment efficiency compared to others (eg: a biconical outflow), there will be preferred sightlines that exhibit high column density and others will exhibit a relatively lower column density. This is illustrated in Fig. \ref{fig:Outflow_geometry}. In addition, projection effects become more important in a non-spherical outflow as opposed to a spherical outflow.
Therefore, we speculate that the increased spread could indicate development of a non-spherical outflow in the high-SFR systems, while the low-SFR galaxies having more uniform/spherical outflows. 

Such development of outflow collimation naturally (hydrodynamically) emerges in the recent TNG simulation results along the minor axis of the galaxy despite isotropic injection of the stellar feedback \citep{nelson2019first}. Thus, they come to a similar conclusion in their simulations: the mass outflow rate of winds is not directionally isotropic, even for $\mathrm{M_*} = 10^{10} \mathrm{M_{\odot}}$ at $z = 1$. While they suggest that the effect is more pronounced as we go below $z \sim 2$ due to the rise of ordered rotation (and resulting emergence of galactic discs), it will be interesting to probe how this effect evolves in the redshift-SFR space. 




\subsection{Evolution with redshift}
While we see that $\mathrm{V_{max}}$ for high-ion lines is correlated with SFR and $\mathrm{M_{*}}$, we do not observe any systematic evolution in $\mathrm{V_{max}}$ in our two redshift groups ($z1: 2-2.7$ and $z2: 2.7 - 6$).  This is consistent with a weak or no evolution ($\mathrm{V_{max}}$  $\sim$ $(1+z)^{0.5}$) suggested in 
\cite{sugahara2017evolution, sugahara2019fast}. However, we do not observe as high $\mathrm{V_{max}}$ values as seen in these studies. It is possible that this discrepancy is due to the small sample size (7 galaxies) in \cite{sugahara2019fast} or due to difference in the observational technique (down-the-barrel $vs$ GRB sightlines). 



\section{Summary}
We explored the outflow-galaxy correlations in low-mass ($\mathrm{M_*}$ $\sim$ $10^9 - 10^{11}$ $\mathrm{M_{\odot}}$) star-forming galaxies at $z \sim 2-6$ using GRB sightline spectroscopy. This technique offers a narrow, off-centered, and randomly oriented sightline as opposed to a broad, down-the-barrel beam to probe the outflows in absorption. We summarize our results as follows:

\noindent
1. SFR-driven outflows: We find strong correlations between outflow column density ($\mathrm{N_{out}}$), outflow maximum velocity ($\mathrm{V_{max}}$) and SFR. This correlation is stronger for high-ion outflows compared to low-ion outflows.

\noindent
2. Correlation with $\mathrm{M_{*}}$: We find a large spread, and therefore, weaker correlations of $\mathrm{N_{out}}$ and $\mathrm{V_{max}}$ in high-ion outflows with stellar mass. We find that this spread is almost entirely driven by the spread in SFR in a given stellar mass bin. On the other hand, we find that low-ion outflows are not correlated with stellar mass. 

\noindent
3. We observe a higher detection fraction in high-ions compared to low-ions (irrespective of the SFR or $\mathrm{M_{*}}$) as well as typically higher $\mathrm{V_{max}}$ and $\mathrm{N_{out}}$ in high-ions, indicating that the outflow is dominated by the phase traced by high-ionization lines. 

\noindent
4. We investigated how the normalized velocity ($\mathrm{V_{norm}}$ = $\mathrm{V_{max}/V_{circ, halo}}$) depends on the halo mass ($\mathrm{M_{halo}}$). We find a decline in $\mathrm{V_{norm}}$ with increasing halo mass and thereby, infer that the outflows from the low-mass halos are more likely to escape and enrich the outer CGM and/or IGM than those in the halos of higher masses.
We also observe a steeper decline in $\mathrm{V_{norm}}$ for low-ions potentially hinting towards a systematic shift in the temperature distribution of the outflow as the halo mass increases. 

\noindent
5. sSFR and metal enrichment: While neither $\mathrm{N_{out}}$ nor $\mathrm{V_{max}}$ are correlated with sSFR, we observe a strong correlation between $\mathrm{V_{norm}}$ (= $\mathrm{V_{max}/V_{circ, halo}}$) and sSFR at the 3-$\sigma$ and 3.7-$\sigma$ levels for C IV and Si IV, respectively, with $\mathrm{V_{norm}}$ $\propto$ sSFR$^{0.25}$. This power law is consistent with the low-redshift results from \cite{heckman2016implications}. From this result, we infer that the outflows in galaxies with higher sSFR have higher velocities relative to the characteristic velocity of their halos. Thus, the outflows from high sSFR galaxies are more likely to escape and enrich the outer CGM and IGM compared to low sSFR galaxies. 

\noindent 
6 Outflow dynamics: Using N$\cdot$V and  N$\cdot$V$^2$ as proxies for $\dot{M}_{out}$ and $\dot{p}_{out}$ respectively,  we find that they are both correlated to the SFR for the high-ion traced outflows. We do not find such a correlation for low-ions. While the high-SFR galaxies have a higher outflow rate, the low-SFR galaxies are more efficient in driving outflows normalized to their SFRs. In addition, we find that the estimated momentum flux of the outflow can easily be supported by the momentum injected by stellar process (i.e. supernovae and stellar winds). This, in concert with the observed correlations, implies that the blue wings in the high-ion absorption are dominated by star-formation-driven outflows.   

\noindent
7. Redshift evolution: We do not observe any systematic difference in the $\mathrm{V_{max}-SFR}$ and $\mathrm{N_{out}-SFR}$ correlations (for high-ions) in our two redshift groups ($z1$: 2$-$2.7 and $z2:$ $2.7-6$) of similar size. This suggests that the correlations of the outflow with SFR do not significantly evolve with redshift and are more fundamental in nature. 

\noindent
8. Stacking comparison: By comparing the stacks of our CGM-GRB sample with the down-the-barrel studies of starbursts at $z \sim 2$ (MEGaSaURA) and at $z \sim 0.1$ (COS-sample) in Section \ref{sec:stacking_comparison},  we find that over a broad redshift range, the strength of the outflow is correlated with the main-sequence offset at the respective redshifts rather than the simple star-formation rate.

\noindent
9. Structure of the outflow: We observe a larger spread in the high-ion correlation with SFR beyond an SFR of $\sim$10 $\mathrm{M_{\odot} yr^{-1}}$. We speculate that this spread could arise due to an emergence of non-spherical outflows (eg: biconical) at high-SFRs, leading to some sightlines cutting across a larger section of the outflow while others probing a weaker outflow. This characteristic can be uniquely probed using the GRB sightline technique due to the random orientation (i.e. not down the barrel) and narrow beam of the sightline (as shown in Fig. \ref{fig:Outflow_geometry}. However, a more detailed investigation of the physical processes in the outflows is needed to explain the observed spread in the outflow column densities at high SFR.

These results highlight the unique potential of GRB afterglow spectroscopy to explore the nature and importance of stellar feedback at high redshifts.\\

The authors are grateful to Drs.\ A. Cucchiara  and V. Toy for their useful comments in the early stages of this paper. 
P.G. was supported by NASA Earth and Space Science Fellowship (ASTRO18F-0085), NASA Hubble Fellowship (HST-HF2-51478.001-A), and David \&  Ellen Lee Fellowship at Caltech for this research. 
S.V. acknowledges partial support from the National Science Foundation under grant 1711377 and National Aeronautics and Space Administration under grants 16-APRA 16-0064 and ADAP NNX16AF24G. 

These results made use of the Lowell Discovery Telescope (LDT) at Lowell Observatory. Lowell is a private, non-profit institution dedicated to astrophysical research and public appreciation of astronomy and operates the LDT in partnership with Boston University, the University of Maryland, the University of Toledo, Northern Arizona University and Yale University. The Large Monolithic Imager was built by Lowell Observatory using funds provided by the National Science Foundation (AST-1005313).

Based on observations made with ESO Telescopes at the La Silla Paranal Observatory under programme IDs 177.A-3016, 177.A-3017, 177.A-3018 and 179.A-2004, and on data products produced by the KiDS consortium. The KiDS production team acknowledges support from: Deutsche Forschungsgemeinschaft, ERC, NOVA and NWO-M grants; Target; the University of Padova, and the University Federico II (Naples). 
 
\appendix

\section{Covering fraction}
A covering fraction ($C_f$) of 1 was assumed in deriving the Voigt-profile fits and column densities used in this paper \cite{gatkine2019cgm}. Here we present a more quantitative rationale for that assumption. A typical feature of fractional covering fraction seen in partial covering fraction cases is line saturation with residual positive flux. We do not see that in any of the GRB sightlines in this sample, suggesting a high covering fraction. All the line saturations go to zero flux  (see Figures 17-43 in \citealt{gatkine2019cgm}).

Secondly, in \cite{gatkine2019cgm}, all the doublets were fit using a joint Bayesian fitting for the doublets with an assumption of $C_f$ = 1 resulting in reasonable fits for both the components of the doublets, which suggest that a $C_f$ of 1 is a consistent assumption.

Lastly, the UV emitting region for a GRB during the first 1-2 days after the GRB explosion (typically the timescale when the spectra are obtained) is extremely tiny ($\sim$ $1-2$ light days $\sim$ 100$-$200 AU). This is much more compact than the typical cloud sizes expected for the outflows which can be estimated as $\mathrm{\frac{total~ column~density}{density}}$ $\sim$ $10^{18}/10^{-2}$ cm $\sim$ 10$^{20}$ cm $>>$ 100-200 AU. Here, we take typical HI column density as the total column density and 10$^{-2}$ cm$^{-3}$ as a typical density for the  outflow absorbers.  Similarly,   \cite{liang2020model} suggest a cloud diameter of 1-2 pc.
Hence, the GRB sightline can be treated as a pencil beam and therefore, a covering fraction of 1 is a reasonable assumption.

\begin{figure*}
\centering
\includegraphics[width=0.85\textwidth]{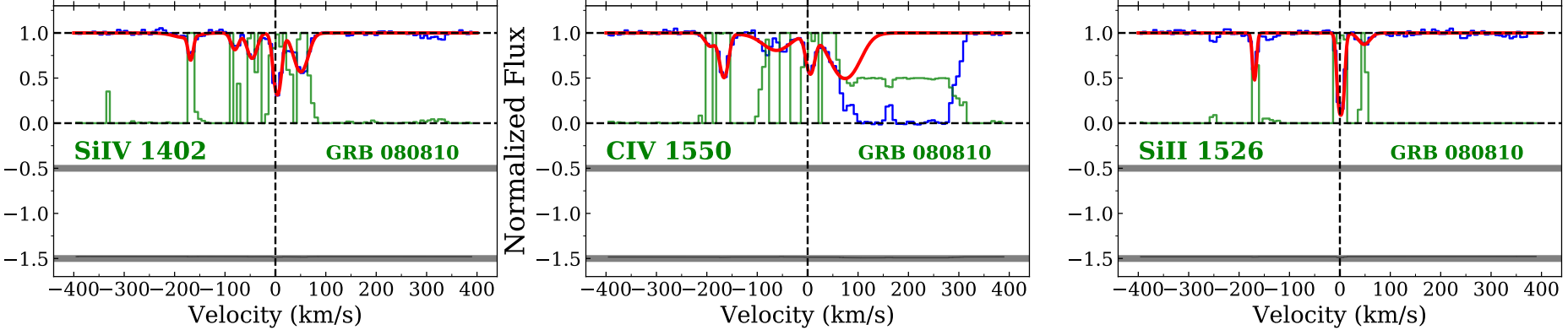}
\figcaption{\label{fig:Covering_Fraction} 
Covering fractions calculated for C IV, Si IV doublets and approximated for Si II 1526 in GRB 080810. The blue, red, and green lines show the normalized flux, Voigt-profle fit, and covering fraction, respectively. The error in the flux  is shown in the bottom panel offset by $-1.5$.}
\end{figure*}

Furthermore, we computed the $C_f$ for the C IV and Si IV doublets, and approximated it for the Si II 1526 $-$ Si II 1190 pair using the formalism described in \cite{hamann1997high}. An example is shown in Fig. \ref{fig:Covering_Fraction}.
Note that the ratio of  (oscillator strength $\times$ $\lambda$) needs to be $\sim$2 for the formalism in \cite{hamann1997high} and hence, the $C_f$ for Si II 1526 is only an approximation.  
The covering fraction, shown in green is always $\gtrsim$ 0.75 for the line components. Therefore, the assumption of $C_f$ = 1 is a reasonable assumption for our CGM-GRB sample probed by the GRB sightline.

\section{Additional outflow-galaxy correlations}\label{app:Additional_correlations}
In this appendix section, we discuss the weaker ($<$ 2$\sigma$) outflow-galaxy correlations to provide the full scope of correlations investigated for the CGM-GRB sample. Deeper observations of the host galaxies and a larger sample will be required to evaluate these correlations in the future.

\subsection{${N_{out}}$ $vs$ ${M_*}$} 
\label{subsubsec:N_vs_M*}
\begin{figure*}
\centering
\includegraphics[width=0.85\textwidth]{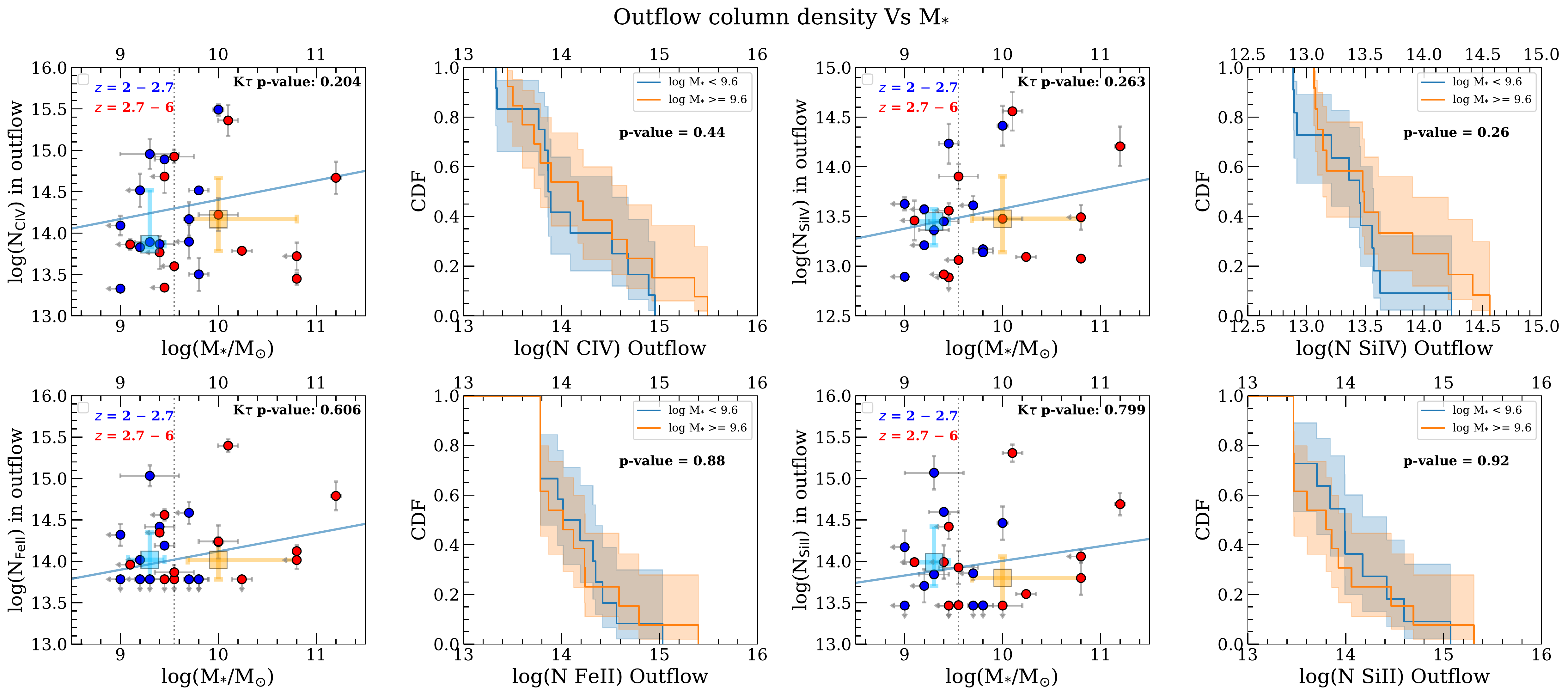}
\figcaption{\label{fig:N_vs_M*} 
Same as Figure \ref{fig:N_vs_SFR}, for the outflow column density $vs$ $\mathrm{M_*}$}
\end{figure*}

From visual inspection of the $\mathrm{N_{out}}$ $vs$ $\mathrm{M_*}$ panels in Fig. \ref{fig:N_vs_M*}, there is a minor rise in $\mathrm{N_{out}}$ of high-ions with stellar mass (better seen in C IV and Si IV), albeit with weak statistical significance (i.e. a high value of Kendall-$\tau$ p-value). More quantitatively, the correlation exists with a confidence level of 80\% and 74\% in C IV and Si IV respectively. 
Si IV plot also shows a significantly larger spread in column density at a higher stellar mass. On the other hand, the low-ions do not show any difference between the low-mass and high-mass samples (as evident from the CDF plots and high $p$-values). Thus, the overall column density of the low-ion outflow remains independent of the stellar mass and shows a weak correlation for the column density of high-ion outflow. This finding implies that the prevalence of low-ion traced outflows is largely independent of the stellar mass and for high-ion outflows (particularly Si IV), it is only moderately boosted at high $\mathrm{M_{*}}$. Given the large spread ($\sim$ 2 dex) in the column densities at any mass (for both high- and low-ions), it is clear that the relation between outflow column density and stellar mass, if any, is complex with multiple contributing factors such as halo mass, SFR, and ionization state.

\subsection{${N_{out}}$ $vs$ sSFR} 
\label{subsubsec:N_vs_sSFR}
\begin{figure*}
\centering
\includegraphics[width=0.95\textwidth]{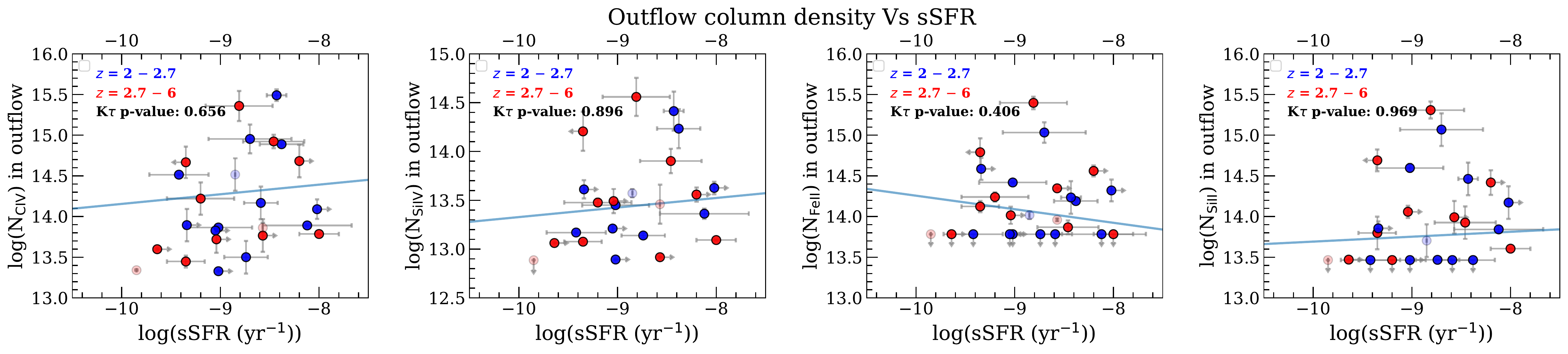}
\figcaption{\label{fig:N_vs_sSFR} 
Same as Figure \ref{fig:N_vs_SFR}, for outflow column density $vs$ specific SFR. The objects where both SFR and $\mathrm{M_*}$ are not detected are shown in a lighter shade.}
\end{figure*}

We investigated the correlations, if any, between outflow column density ($\mathrm{N_{out}}$) and specific SFR (sSFR = SFR/$\mathrm{M_{*}}$) in Fig. \ref{fig:N_vs_sSFR}. We do not see any statistically significant correlation. 
Previous studies at lower redshifts have shown only a weak or no correlation between outflow column density and sSFR for either low-ion or high-ion outflows. For instance, \cite{du2016kinematics} see only a weak correlation for C IV-traced outflows at $z \sim 1.25$ while \cite{bradshaw2013high} see no correlation for Mg II-traced outflows at $z \sim 0.7 - 1.63$. Thus, the lack of $\mathrm{N_{out}-sSFR}$ relations for both high and low ions at $z \sim 2-6$ in our data, are consistent with previous results at lower redshifts.

\subsection{Outflow $\mathrm{V_{max}}$ $vs$ $\mathrm{M_*}$}
\begin{figure*}
\centering
\includegraphics[width=0.85\textwidth]{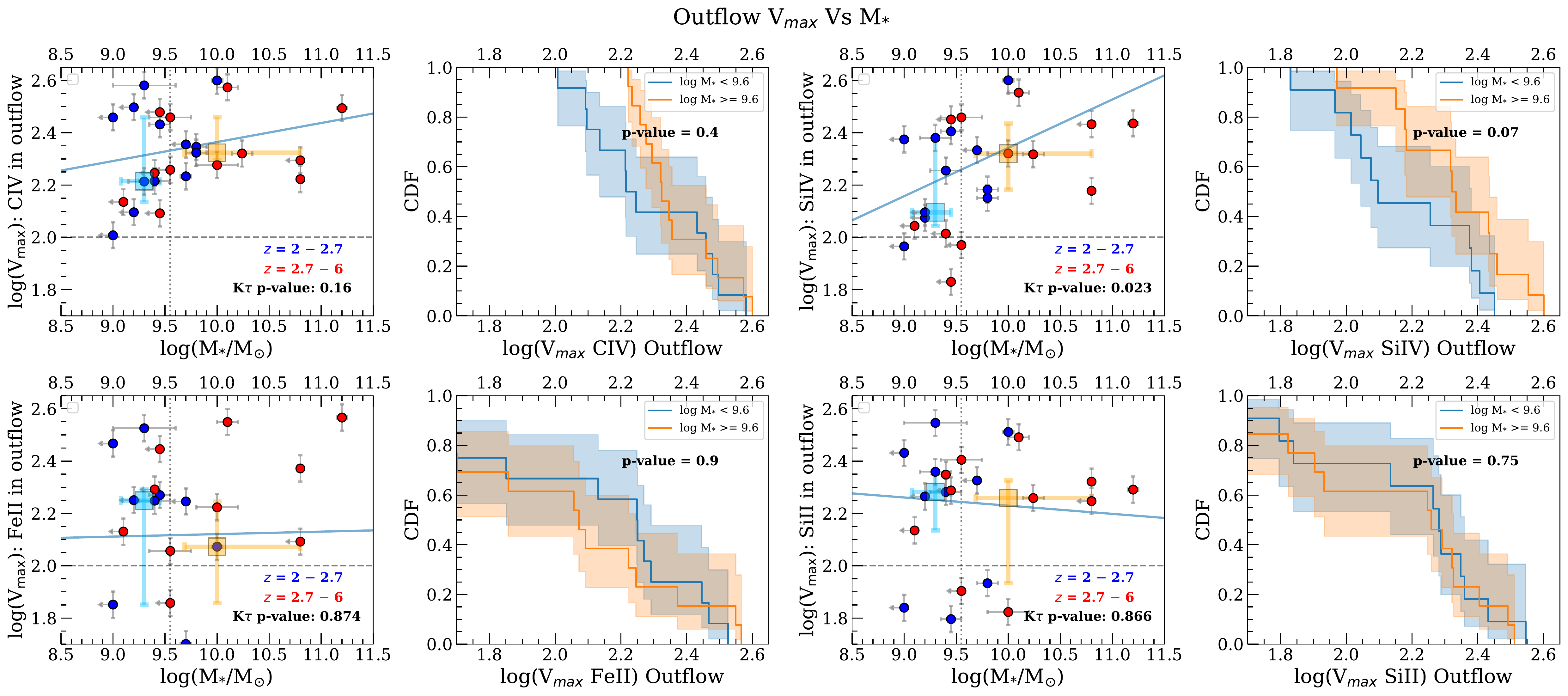}
\figcaption{\label{fig:Vmax_vs_M*} 
Same as Figure \ref{fig:N_vs_SFR}, for the maximum outflow velocity, $\mathrm{V_{max}}$ $vs$ $\mathrm{M_{*}}$. The horizontal dashed line in the panels shows the 100 km s$^{-1}$ level, which we treat as the threshold for outflow. }
\end{figure*}

This relation is summarized in Fig. \ref{fig:Vmax_vs_M*}. We observe a weak correlation in C IV (84\% confidence or $\sim$ 1.5$\sigma$) and a slightly stronger correlation in Si IV (98\% confidence or $\sim$ 2.5$\sigma$). We note a larger spread in the low-$\mathrm{M_{*}}$ group, as evident from the errorbars around the low-$\mathrm{M_{*}}$ in Fig. \ref{fig:Vmax_vs_M*} (top panels). This spread can be directly explained by the larger spread of SFR in the low-$\mathrm{M_{*}}$ group  compared to the high-$\mathrm{M_{*}}$ group in Fig. \ref{fig:Sample_properties} (see panel 2). It is the variance in SFR that is directly causing the spread in the  $\mathrm{V_{max}}-\mathrm{M_*}$ plot for high ions. By combining this with the result from Section \ref{subsubsec:N_vs_M*}, we can deduce that the stellar mass by itself does not significantly affect the kinematics or column density of the high-ion outflows, but is almost entirely driven by the SFR instead. 

For low-ion species, we do not see any correlation between $\mathrm{V_{max}}$ and $\mathrm{M_*}$. We observe a large spread in both high-mass and low-mass groups. Overall, our high-$z$ results for the low-ion traced outflows show a different picture compared to the results in low-$z$ studies. For instance, \cite{rubin2014evidence} find a 3.5-$\sigma$ correlation between $\mathrm{V_{max}}$ and $\mathrm{M_*}$ and no correlation with current SFR. In contrast, for our sample at $z > 2$, we find a weak correlation ($\sim$ 2$\sigma$ for Si II) with SFR and no correlation with the stellar mass. 

\subsection{Outflow $\mathrm{V_{max}}$ $vs$ sSFR}
\label{subsubsec:Vmax_vs_sSFR}
\begin{figure*}
\centering
\includegraphics[width=0.95\textwidth]{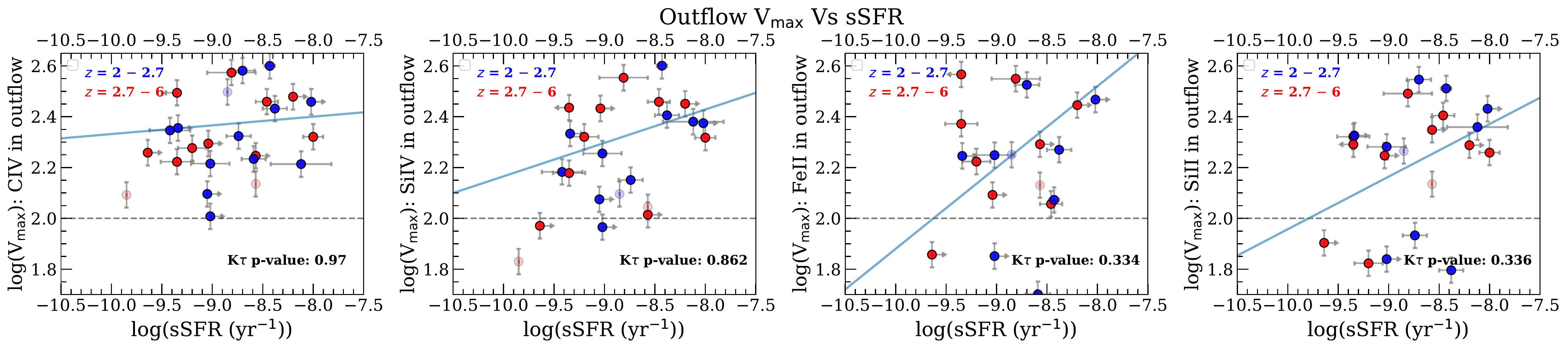}
\figcaption{\label{fig:Vmax_vs_sSFR} 
Same as Figure \ref{fig:N_vs_sSFR}, for the the scaling relations of the maximum outflow velocity ($\mathrm{V_{max}}$) with specific SFR (= SFR / $\mathrm{M_*}$). The horizontal dashed line in the panels shows the 100 km s$^{-1}$ level, which we treat as the threshold for outflow.}
\end{figure*}

We summarize the results of $\mathrm{V_{max}}$ $vs$ sSFR correlation 
in Fig. \ref{fig:Vmax_vs_sSFR}. We do not find any statistically significant correlation of sSFR with $\mathrm{V_{max}}$. A weak, 1-$\sigma$ correlation appears to be present for $\mathrm{V_{max}}$ $vs$ sSFR for low-ion outflows, albeit with large scatter. 

Correlations between outflow kinematics and sSFR have been seen in past observations. For instance, \cite{heckman2015systematic, heckman2016implications} report a strong ($>$ 2 $\sigma$) correlation between outflow velocity of warm ionized gas and sSFR over 2.5 orders of magnitude in sSFR for starburst galaxies at $z < 0.2$. \cite{bradshaw2013high} suggest a $\mathrm{V_{max}-sSFR}$ correlation at $z \sim 0.7 - 1.63$, but shallower compared to $z \sim 0$ correlation. 
\cite{sugahara2017evolution} also report $\mathrm{V_{max}-sSFR}$ correlations at $z \sim 0-1$, albeit with a significant variation with redshift and over only one order of magnitude in sSFR. For comparison, we explore this correlation over almost 2 orders of magnitude in sSFR and do not find any statistically significant correlation. 

This lack of correlation maybe due to a number of reasons: 1) an intrinsic weakening of the correlation at high redshifts, 2) presence of a large scatter in the correlation, thus requiring observations over a larger range in sSFR to see the correlation, and 3) the upper and lower limits in sSFR may mask the underlying correlation. In all the three cases, deeper observations of the host galaxy are needed to better constrain the sSFR and thereby, its relation with the outflow properties.   

\section{Consideration of other low-ion lines: C II 1334 and O I 1302}\label{app:low-ions}
\begin{figure*}
\centering
\includegraphics[width=0.85\textwidth]{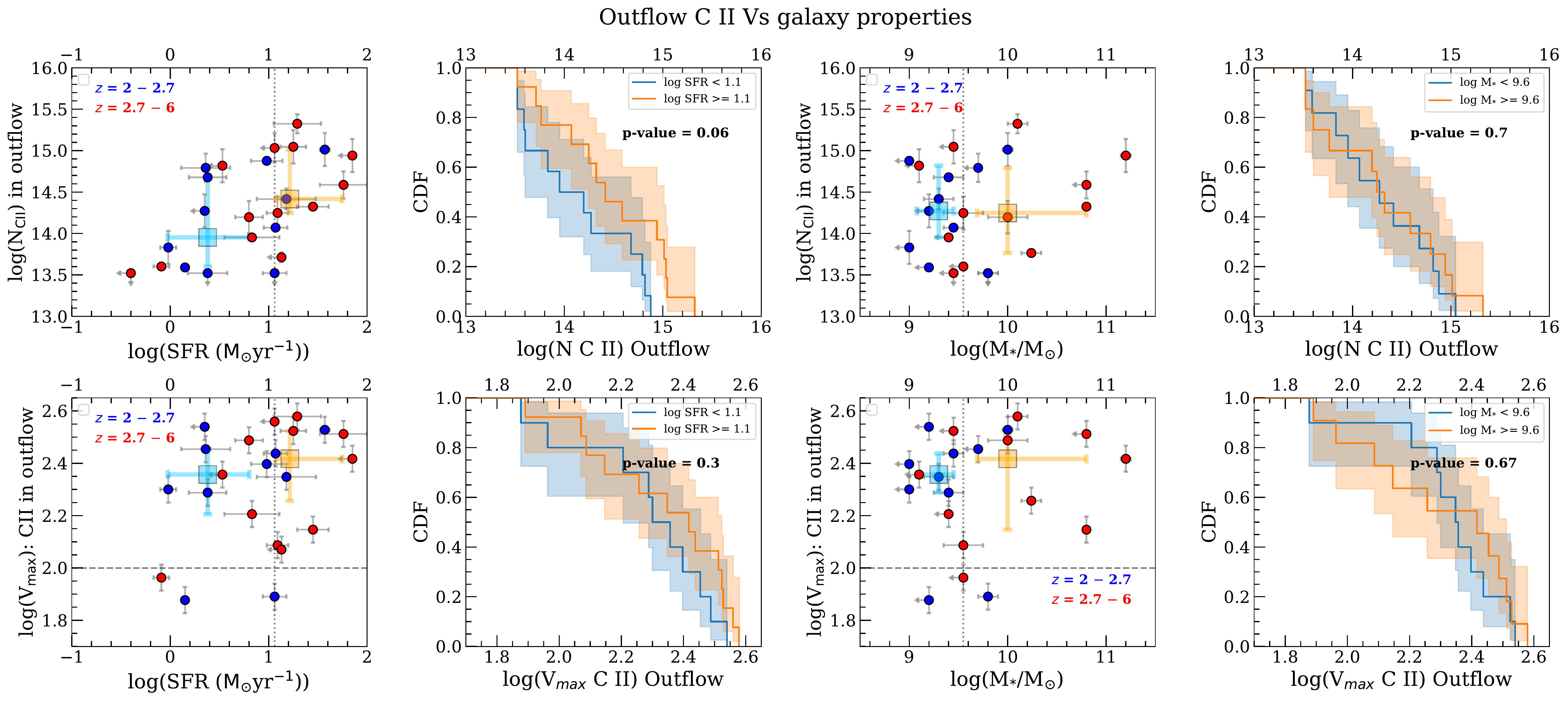}
\figcaption{\label{fig:CII_relations} 
Same as Figure \ref{fig:N_vs_SFR}, for the correlations of the C II 1334-traced outflow with galaxy properties.}
\end{figure*}

\begin{figure*}
\centering
\includegraphics[width=0.85\textwidth]{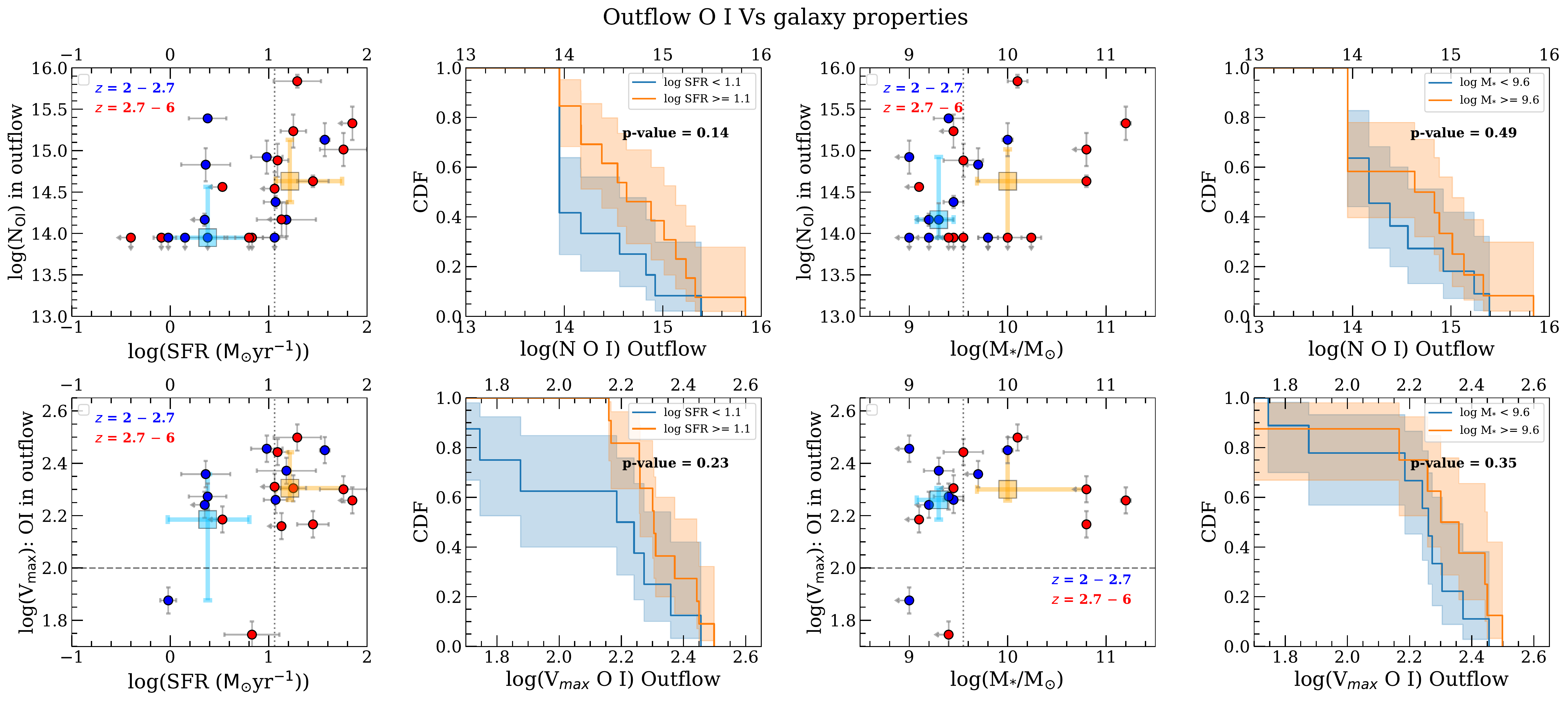}
\figcaption{\label{fig:OI_relations} 
Same as Figure \ref{fig:N_vs_SFR}, for the correlations of the O I 1302-traced outflow with galaxy properties.}
\end{figure*}

As previously explained in section \ref{sec:blue_wing_column}, we did not use C II 1334 and O I 1302 for the primary investigation of the outflow-galaxy correlations due to mild concerns about possible blending due to neighboring lines. However, we performed the same analysis for these two lines as a consistency check for the low-ion correlations  observed in the main paper. Here we mainly focus on $\mathrm{N_{out}}$ and $\mathrm{V_{max}}$ $vs$ $\mathrm{M_*}$ and SFR. These plots are show in Figs.  \ref{fig:CII_relations} and \ref{fig:OI_relations}. We observe that the $\mathrm{N_{out, C II}}$ $-$ SFR correlation has a smaller p-value ($\sim$ $2\sigma$) compared to that observed for other low-ion lines. Other than this deviation, we do not see any significant departure in the results compared to those obtained for Si II 1526 and Fe II 1608.   

\section{Relation of O VI absorption with galaxy properties}
O VI traces the warm-hot medium ($\mathrm{10^5 - 10^6}$ K, assuming collisional equilibrium). It is likely that O VI absorption traces a different phase than the gas traced by C IV and Si IV. Here we look at the same relations studied above with O VI. Due to the location of the O VI doublet in the far UV ($\lambda_{rest}$ = 1031.9 and 1037.6 \AA), fewer afterglow spectra in our sample cover the O VI band. Therefore, we caution that the relations observed for O VI are based on a smaller sample than other lines. Regardless, O VI correlations can provide useful insights into the warm-hot phase of the CGM. 

\begin{figure*}
\centering
\includegraphics[width=0.85\textwidth]{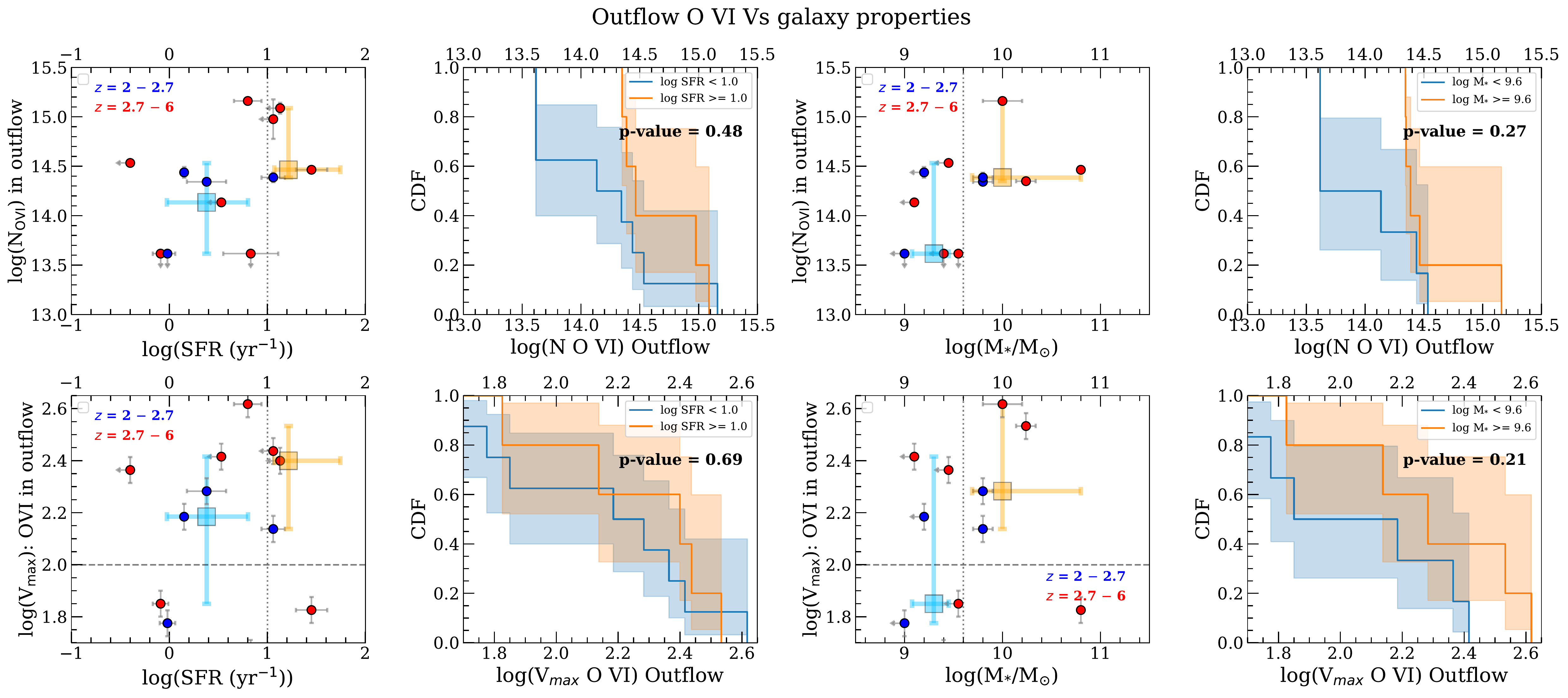}
\figcaption{\label{fig:OVI_relations} 
Same as Figure \ref{fig:N_vs_SFR}, for the correlations of the O VI-traced outflow with galaxy properties.}
\end{figure*}

From Fig. \ref{fig:OVI_relations}, it is clear that both the outflow column density and $\mathrm{V_{max}}$ correlate with the stellar mass. The outflow column density is also correlated to the SFR, albeit less so than with the stellar mass. The O VI correlations appear to follow the correlations of high-ion species (C IV, Si IV) described earlier. With a caution of limited sample size, the data indicates that star formation activity also drives outflows in the OVI-traced phase at $z \sim 2-6$. 



\begin{deluxetable}{lcccccc}[t]
\tablecaption{Summary of correlations between CGM  and galaxy properties investigated for this sample \label{tab:CGM_GRB_correlations}}
\tablehead{
\colhead{Outflow Property} &
\colhead{Galaxy Property} &
\colhead{Ion} &
\colhead{Slope} & 
\colhead{Intercept} &
\colhead{K$-\tau$ p-value} &
\colhead{Median-split p-value} 
}

\startdata
{\color[HTML]{101010} Outflow Column Density (N$\mathrm{_{out}}$)}                             & {\color[HTML]{101010} M$_*$}    & {\color[HTML]{101010} C IV}  & {\color[HTML]{101010} 0.23}  & {\color[HTML]{101010} 12.08} & {\color[HTML]{101010} 0.20} & {\color[HTML]{101010} 0.44}                  \\
{\color[HTML]{101010} Outflow Column Density (N$\mathrm{_{out}}$)}                             & {\color[HTML]{101010} M$_*$}    & {\color[HTML]{101010} Si IV} & {\color[HTML]{101010} 0.20}  & {\color[HTML]{101010} 11.57} & {\color[HTML]{101010} 0.26} & {\color[HTML]{101010} 0.26}                  \\
{\color[HTML]{101010} Outflow Column Density (N$\mathrm{_{out}}$)}                             & {\color[HTML]{101010} M$_*$}    & {\color[HTML]{101010} Fe II} & {\color[HTML]{101010} 0.22}  & {\color[HTML]{101010} 11.91} & 
{\color[HTML]{101010} 0.61} & {\color[HTML]{101010} 0.88}                  \\
{\color[HTML]{101010} Outflow Column Density (N$\mathrm{_{out}}$)}                             & {\color[HTML]{101010} M$_*$}    & {\color[HTML]{101010} Si II} & {\color[HTML]{101010} 0.18}  & {\color[HTML]{101010} 12.24} & {\color[HTML]{101010} 0.80} & {\color[HTML]{101010} 0.92}                  \\
\hline
{\color[HTML]{101010} Outflow Column Density (N$\mathrm{_{out}}$)}                             & {\color[HTML]{101010} SFR}     & {\color[HTML]{101010} C IV}  & {\color[HTML]{101010} 0.22}   & {\color[HTML]{101010} 13.99} & {\color[HTML]{101010} 0.14} & {\color[HTML]{101010} 0.13}                  \\
{\color[HTML]{101010} Outflow Column Density (N$\mathrm{_{out}}$)}                             & {\color[HTML]{101010} SFR}     & {\color[HTML]{101010} Si IV} & {\color[HTML]{101010} 0.25}  & {\color[HTML]{101010} 13.25} & {\color[HTML]{101010} 0.08} & {\color[HTML]{101010} 0.02}                  \\
{\color[HTML]{101010} Outflow Column Density (N$\mathrm{_{out}}$)}                             & {\color[HTML]{101010} SFR}     & {\color[HTML]{101010} Fe II} & {\color[HTML]{101010} 0.12}   & {\color[HTML]{101010} 13.86} & {\color[HTML]{101010} 0.53} & {\color[HTML]{101010} 0.98}                  \\
{\color[HTML]{101010} Outflow Column Density (N$\mathrm{_{out}}$)}                             & {\color[HTML]{101010} SFR}     & {\color[HTML]{101010} Si II} & {\color[HTML]{101010} 0.27}   & {\color[HTML]{101010} 13.65} & {\color[HTML]{101010} 0.20} & {\color[HTML]{101010} 0.53}                  \\
\hline
{\color[HTML]{101010} Maximum outflow velocity (V$\mathrm{_{max}}$)}                  & {\color[HTML]{101010} M$_*$}    & {\color[HTML]{101010} C IV}  & {\color[HTML]{101010} 0.07}  & {\color[HTML]{101010} 1.64}  & {\color[HTML]{101010} 0.16} & {\color[HTML]{101010} 0.4}                   \\
{\color[HTML]{101010} Maximum outflow velocity (V$\mathrm{_{max}}$)}                  & {\color[HTML]{101010} M$_*$}    & {\color[HTML]{101010} Si IV} & {\color[HTML]{101010} 0.18}  & {\color[HTML]{101010} 0.49}  & {\color[HTML]{101010} 0.02}  & {\color[HTML]{101010} 0.07}                  \\
{\color[HTML]{101010} Maximum outflow velocity (V$\mathrm{_{max}}$)}                  & {\color[HTML]{101010} M$_*$}    & {\color[HTML]{101010} Fe II} & {\color[HTML]{101010} 0.0095}  & {\color[HTML]{101010} 2.03}  & {\color[HTML]{101010} 0.87} & {\color[HTML]{101010} 0.9}                   \\
{\color[HTML]{101010} Maximum outflow velocity (V$\mathrm{_{max}}$)}                  & {\color[HTML]{101010} M$_*$}    & {\color[HTML]{101010} Si II} & {\color[HTML]{101010} -0.03} & {\color[HTML]{101010} 2.54}  & {\color[HTML]{101010} 0.87} & {\color[HTML]{101010} 0.75}                  \\
\hline
{\color[HTML]{101010} Maximum outflow velocity (V$\mathrm{_{max}}$)}                  & {\color[HTML]{101010} SFR}     & {\color[HTML]{101010} C IV}  & {\color[HTML]{101010} 0.12}  & {\color[HTML]{101010} 2.19}  & {\color[HTML]{101010} 0.05}  & {\color[HTML]{101010} 0.33}                  \\
{\color[HTML]{101010} Maximum outflow velocity (V$\mathrm{_{max}}$)}                  & {\color[HTML]{101010} SFR}     & {\color[HTML]{101010} Si IV} & {\color[HTML]{101010} 0.28}  & {\color[HTML]{101010} 2.07}  & {\color[HTML]{101010} 0.0007} & {\color[HTML]{101010} 0.0001}                \\
{\color[HTML]{101010} Maximum outflow velocity (V$\mathrm{_{max}}$)}                  & {\color[HTML]{101010} SFR}     & {\color[HTML]{101010} Fe II} & {\color[HTML]{101010} 0.06}  & {\color[HTML]{101010} 2.08}  & {\color[HTML]{101010} 0.61} & {\color[HTML]{101010} 0.66}                  \\
{\color[HTML]{101010} Maximum outflow velocity (V$\mathrm{_{max}}$)}                  & {\color[HTML]{101010} SFR}     & {\color[HTML]{101010} Si II} & {\color[HTML]{101010} 0.14}  & {\color[HTML]{101010} 2.10}  & {\color[HTML]{101010} 0.07} & {\color[HTML]{101010} 0.82}                  \\
\hline
{\color[HTML]{101010} Normalized Velocity (V$\mathrm{_{max}}$/V$\mathrm{_{circ, halo}}$)}        & {\color[HTML]{101010} M$\mathrm{_{halo}}$} & {\color[HTML]{101010} C IV}  & {\color[HTML]{101010} -0.32} & {\color[HTML]{101010} 3.92}  & {\color[HTML]{101010} 0.003}  & {\color[HTML]{101010} 0.24}                  \\
{\color[HTML]{101010} Normalized Velocity (V$\mathrm{_{max}}$/V$\mathrm{_{circ, halo}}$)}        & {\color[HTML]{101010} M$\mathrm{_{halo}}$} & {\color[HTML]{101010} Si IV} & {\color[HTML]{101010} -0.45} & {\color[HTML]{101010} 5.34}  & {\color[HTML]{101010} 0.0012} & {\color[HTML]{101010} 0.98}                  \\
{\color[HTML]{101010} Normalized Velocity (V$\mathrm{_{max}}$/V$\mathrm{_{circ, halo}}$)}        & {\color[HTML]{101010} M$\mathrm{_{halo}}$} & {\color[HTML]{101010} Fe II} & {\color[HTML]{101010} -0.73} & {\color[HTML]{101010} 8.34}  & {\color[HTML]{101010} 0.14} & {\color[HTML]{101010} 0.12}                  \\
{\color[HTML]{101010} Normalized Velocity (V$\mathrm{_{max}}$/V$\mathrm{_{circ, halo}}$)}        & {\color[HTML]{101010} M$\mathrm{_{halo}}$} & {\color[HTML]{101010} Si II} & {\color[HTML]{101010} -0.51}  & {\color[HTML]{101010} 5.85}  & {\color[HTML]{101010} 0.0085} & {\color[HTML]{101010} 0.04}                  \\
\hline
{\color[HTML]{101010} Outflow Column Density (N$\mathrm{_{out}}$)}                             & {\color[HTML]{101010} sSFR}    & {\color[HTML]{101010} C IV}  & {\color[HTML]{101010} 0.12}  & {\color[HTML]{101010} 15.34} & {\color[HTML]{101010} 0.66} & \multicolumn{1}{l}{{\color[HTML]{101010} -}} \\
{\color[HTML]{101010} Outflow Column Density (N$\mathrm{_{out}}$)}                             & {\color[HTML]{101010} sSFR}    & {\color[HTML]{101010} Si IV} & {\color[HTML]{101010} 0.098}  & {\color[HTML]{101010} 14.31} & {\color[HTML]{101010} 0.90} & \multicolumn{1}{l}{{\color[HTML]{101010} -}} \\
{\color[HTML]{101010} Outflow Column Density (N$\mathrm{_{out}}$)}                             & {\color[HTML]{101010} sSFR}    & {\color[HTML]{101010} Fe II} & {\color[HTML]{101010} -0.17} & {\color[HTML]{101010} 12.59} & {\color[HTML]{101010} 0.41} & \multicolumn{1}{l}{{\color[HTML]{101010} -}} \\
{\color[HTML]{101010} Outflow Column Density (N$\mathrm{_{out}}$)}                             & {\color[HTML]{101010} sSFR}    & {\color[HTML]{101010} Si II} & {\color[HTML]{101010} 0.06}  & {\color[HTML]{101010} 14.30} & {\color[HTML]{101010} 0.97} & \multicolumn{1}{l}{{\color[HTML]{101010} -}} \\
\hline
{\color[HTML]{101010} Maximum outflow velocity (V\_max)}                  & {\color[HTML]{101010} sSFR}    & {\color[HTML]{101010} C IV}  & {\color[HTML]{101010} 0.03}  & {\color[HTML]{101010} 2.67}  & {\color[HTML]{101010} 0.97} & \multicolumn{1}{l}{{\color[HTML]{101010} -}} \\
{\color[HTML]{101010} Maximum outflow velocity (V\_max)}                  & {\color[HTML]{101010} sSFR}    & {\color[HTML]{101010} Si IV} & {\color[HTML]{101010} 0.13}  & {\color[HTML]{101010} 3.48}  & {\color[HTML]{101010} 0.86} & \multicolumn{1}{l}{{\color[HTML]{101010} -}} \\
{\color[HTML]{101010} Maximum outflow velocity (V\_max)}                  & {\color[HTML]{101010} sSFR}    & {\color[HTML]{101010} Fe II} & {\color[HTML]{101010} 0.32}    & {\color[HTML]{101010} 5.08}    & {\color[HTML]{101010} 0.33} & \multicolumn{1}{l}{{\color[HTML]{101010} -}} \\
{\color[HTML]{101010} Maximum outflow velocity (V\_max)}                  & {\color[HTML]{101010} sSFR}    & {\color[HTML]{101010} Si II} & {\color[HTML]{101010} 0.21}   & {\color[HTML]{101010} 4.03}  & {\color[HTML]{101010} 0.34} & \multicolumn{1}{l}{{\color[HTML]{101010} -}} \\
\hline
{\color[HTML]{101010} Normalized Velocity (V$\mathrm{_{max}}$/V$\mathrm{_{circ, halo}}$)}        & {\color[HTML]{101010} sSFR}    & {\color[HTML]{101010} C IV}  & {\color[HTML]{101010} 0.26}  & {\color[HTML]{101010} 2.42}  & {\color[HTML]{101010} 0.003} & \multicolumn{1}{l}{{\color[HTML]{101010} -}} \\
{\color[HTML]{101010} Normalized Velocity (V$\mathrm{_{max}}$/V$\mathrm{_{circ, halo}}$)}        & {\color[HTML]{101010} sSFR}    & {\color[HTML]{101010} Si IV} & {\color[HTML]{101010} 0.24}   & {\color[HTML]{101010} 2.19}  & {\color[HTML]{101010} 0.0002} & \multicolumn{1}{l}{{\color[HTML]{101010} -}} \\
{\color[HTML]{101010} Normalized Velocity (V$\mathrm{_{max}}$/V$\mathrm{_{circ, halo}}$)}        & {\color[HTML]{101010} sSFR}    & {\color[HTML]{101010} Fe II} & {\color[HTML]{101010} 0.18}  & {\color[HTML]{101010} 1.38}  & {\color[HTML]{101010} 0.51} & \multicolumn{1}{l}{{\color[HTML]{101010} -}} \\
{\color[HTML]{101010} Normalized Velocity (V$\mathrm{_{max}}$/V$\mathrm{_{circ, halo}}$)}        & {\color[HTML]{101010} sSFR}    & {\color[HTML]{101010} Si II} & {\color[HTML]{101010} 0.39}  & {\color[HTML]{101010} 3.32}  & {\color[HTML]{101010} 0.010} & \multicolumn{1}{l}{{\color[HTML]{101010} -}} \\
\hline
{\color[HTML]{101010} Outflow rate (N·V = $\int N_{a}vdv$)}               & {\color[HTML]{101010} SFR}     & {\color[HTML]{101010} C IV}  & {\color[HTML]{101010} 0.41}  & {\color[HTML]{101010} 15.81} & {\color[HTML]{101010} 0.12} & {\color[HTML]{101010} 0.12}                  \\
{\color[HTML]{101010} Outflow rate (N·V = $\int N_{a}vdv$)}               & {\color[HTML]{101010} SFR}     & {\color[HTML]{101010} Si IV} & {\color[HTML]{101010} 0.74}  & {\color[HTML]{101010} 14.80} & {\color[HTML]{101010} 0.032} & {\color[HTML]{101010} 0.01}                  \\
{\color[HTML]{101010} Outflow rate (N·V = $\int N_{a}vdv$)}               & {\color[HTML]{101010} SFR}     & {\color[HTML]{101010} Fe II} & {\color[HTML]{101010} 0.90}  & {\color[HTML]{101010} 14.40} & {\color[HTML]{101010} 0.48} & {\color[HTML]{101010} 0.91}                  \\
{\color[HTML]{101010} Outflow rate (N·V = $\int N_{a}vdv$)}               & {\color[HTML]{101010} SFR}     & {\color[HTML]{101010} Si II} & {\color[HTML]{101010} 0.31}  & {\color[HTML]{101010} 14.96} & {\color[HTML]{101010} 0.46} & {\color[HTML]{101010} 0.77}                  \\
\hline
{\color[HTML]{101010} Mass loading factor (N·V/SFR = $\int N_{a}vdv$/SFR)} & {\color[HTML]{101010} SFR}     & {\color[HTML]{101010} C IV}  & {\color[HTML]{101010} 0.39}  & {\color[HTML]{101010} 14.39} & {\color[HTML]{101010} 0.94} & {\color[HTML]{101010} 0.42}                  \\
{\color[HTML]{101010} Mass loading factor (N·V/SFR = $\int N_{a}vdv$/SFR)} & {\color[HTML]{101010} SFR}     & {\color[HTML]{101010} Si IV} & {\color[HTML]{101010} 0.59}  & {\color[HTML]{101010} 13.66}   & {\color[HTML]{101010} 0.79} & {\color[HTML]{101010} 0.82}                  \\
{\color[HTML]{101010} Mass loading factor (N·V/SFR = $\int N_{a}vdv$/SFR)} & {\color[HTML]{101010} SFR}     & {\color[HTML]{101010} Fe II} & {\color[HTML]{101010} 1.47}  & {\color[HTML]{101010} 12.31} & {\color[HTML]{101010} 0.77} & {\color[HTML]{101010} 0.04}                  \\
{\color[HTML]{101010} Mass loading factor (N·V/SFR = $\int N_{a}vdv$/SFR)} & {\color[HTML]{101010} SFR}     & {\color[HTML]{101010} Si II} & {\color[HTML]{101010} 0.94}  & {\color[HTML]{101010} 12.94}  & {\color[HTML]{101010} 0.78} & {\color[HTML]{101010} 0.05}                  \\
\hline
{\color[HTML]{101010} Mass loading factor (N·V/SFR = $\int N_{a}vdv$/SFR)} & {\color[HTML]{101010} M$\mathrm{_{halo}}$} & {\color[HTML]{101010} C IV}  & {\color[HTML]{101010} -0.45} & {\color[HTML]{101010} 20.33} & {\color[HTML]{101010} 0.80}  & {\color[HTML]{101010} 0.75}                  \\
{\color[HTML]{101010} Mass loading factor (N·V/SFR = $\int N_{a}vdv$/SFR)} & {\color[HTML]{101010} M$\mathrm{_{halo}}$} & {\color[HTML]{101010} Si IV} & {\color[HTML]{101010} 0.05}  & {\color[HTML]{101010} 13.61} & {\color[HTML]{101010} 0.87} & {\color[HTML]{101010} 0.33}                  \\
{\color[HTML]{101010} Mass loading factor (N·V/SFR = $\int N_{a}vdv$/SFR)} & {\color[HTML]{101010} M$\mathrm{_{halo}}$} & {\color[HTML]{101010} Fe II} & {\color[HTML]{101010} 1.47}  & {\color[HTML]{101010} 12.31} & {\color[HTML]{101010} 0.77} & {\color[HTML]{101010} 0.39}                  \\
{\color[HTML]{101010} Mass loading factor (N·V/SFR = $\int N_{a}vdv$/SFR)} & {\color[HTML]{101010} M$\mathrm{_{halo}}$} & {\color[HTML]{101010} Si II} & {\color[HTML]{101010} 0.94}  & {\color[HTML]{101010} 12.94}  & {\color[HTML]{101010} 0.78} & {\color[HTML]{101010} 0.07}                  \\
\hline
{\color[HTML]{101010} Momentum flux (N·V$^2$= $\int N_{a}v^{2}dv$)}             & {\color[HTML]{101010} SFR}     & {\color[HTML]{101010} C IV}  & {\color[HTML]{101010} 0.32}   & {\color[HTML]{101010} 18.11} & {\color[HTML]{101010} 0.074} & {\color[HTML]{101010} 0.13}                  \\
{\color[HTML]{101010} Momentum flux (N·V$^2$ = $\int N_{a}v^{2}dv$)}             & {\color[HTML]{101010} SFR}     & {\color[HTML]{101010} Si IV} & {\color[HTML]{101010} 0.91}  & {\color[HTML]{101010} 16.90}  & {\color[HTML]{101010} 0.017} & {\color[HTML]{101010} 0.001}                 \\
{\color[HTML]{101010} Momentum flux (N·V$^2$ = $\int N_{a}v^{2}dv$)}             & {\color[HTML]{101010} SFR}     & {\color[HTML]{101010} Fe II} & {\color[HTML]{101010} 0.97}  & {\color[HTML]{101010} 16.54} & {\color[HTML]{101010} 0.46} & {\color[HTML]{101010} 0.96}                  \\
{\color[HTML]{101010} Momentum flux (N·V$^2$ = $\int N_{a}v^{2}dv$)}             & {\color[HTML]{101010} SFR}     & {\color[HTML]{101010} Si II} & {\color[HTML]{101010} 0.25}  & {\color[HTML]{101010} 17.39} & {\color[HTML]{101010} 0.48} & {\color[HTML]{101010} 0.81}                  \\
\hline
{\color[HTML]{101010} Outflow Column Density (N$\mathrm{_{out}}$)}                             & {\color[HTML]{101010} M$_*$}    & {\color[HTML]{101010} O VI}  & {\color[HTML]{101010} 0.37}  & {\color[HTML]{101010} 10.74} & {\color[HTML]{101010} 0.099} & {\color[HTML]{101010} 0.48}                  \\
{\color[HTML]{101010} Outflow Column Density (N$\mathrm{_{out}}$)}                             & {\color[HTML]{101010} SFR}     & {\color[HTML]{101010} O VI}  & {\color[HTML]{101010} 0.06}  & {\color[HTML]{101010} 14.02} & {\color[HTML]{101010} 0.70} & {\color[HTML]{101010} 0.27}                  \\
{\color[HTML]{101010} Maximum outflow velocity (V$\mathrm{_{max}}$)}                  & {\color[HTML]{101010} M$_*$}    & {\color[HTML]{101010} O VI}  & {\color[HTML]{101010} 0.06}  & {\color[HTML]{101010} 1.77}  & {\color[HTML]{101010} 0.35}  & {\color[HTML]{101010} 0.69}                  \\
{\color[HTML]{101010} Maximum outflow velocity (V$\mathrm{_{max}}$)}                  & {\color[HTML]{101010} SFR}     & {\color[HTML]{101010} O VI}  & {\color[HTML]{101010} -0.06} & {\color[HTML]{101010} 2.30}  & {\color[HTML]{101010} 0.94} & {\color[HTML]{101010} 0.21}          
\enddata

\tablenotetext{a}{}
\end{deluxetable}


\vspace{2mm}
\facilities{Very Large Telescopes, Lowell Discovery Telescope, Spitzer Space Telescope, Hubble Space Telescope, Neil Gehrels Swift Observatory}


\software{ {\fontfamily{qcr}\selectfont galfit} \citep{peng2002detailed}, {\fontfamily{astropy}\selectfont astropy} \citep{robitaille2013astropy},  {\fontfamily{qcr}\selectfont ESOreflex} \citep{freudling2013automated},   {\fontfamily{qcr}\selectfont lifelines} \citep{davidson2019lifelines}
}

\bibliographystyle{apj}
\bibliography{ref}

\end{document}